\documentclass[twocolumn, aps, pra, amsmath, amssymb, nofootinbib, superscriptaddress, longbibliography, floatfix, table-of-contents, eqsecnum, dblfloatfix]{revtex4-2}

\usepackage[pdftex]{graphicx}
\usepackage{mathrsfs}
\usepackage[colorlinks, breaklinks, urlcolor={blue}, linkcolor={red}, citecolor={blue}]{hyperref}
\usepackage{array}
\usepackage{amsmath}
\usepackage{type1cm}
\usepackage{lettrine}
\usepackage[english]{babel}
\usepackage{lmodern}
\usepackage{bm}
\usepackage{microtype}
\usepackage{booktabs}
\usepackage[T1]{fontenc}
\usepackage[boxed, vlined]{algorithm2e}
\usepackage{caption}
\usepackage{braket}

\begin{document}

\title{Dynamical Crossover from Markovian to Non-Markovian dynamics  in the strong coupling regime}

\author{Md.~Manirul Ali}
\email{manirul@citchennai.net}
\affiliation{Centre for Quantum Science and Technology, Chennai Institute of Technology, Chennai 600069, India}

\author{Chandrashekar Radhakrishnan}
\email{chandrashekar10@gmail.com}
\affiliation{Centre for Quantum Information, Communication and Computing, Indian Institute of Technology Madras, Chennai 600036, India}
\affiliation{Centre for Quantum Science and Technology, Chennai Institute of Technology, Chennai 600069, India}

\date{\today}

\begin{abstract}
The transient dynamics of quantum coherence of Gaussian states are investigated. The state is coupled to an external environment
which can be described by a Fano-Anderson type Hamiltonian. Solving the quantum Langevin equation, we obtain the Greens functions 
which are used to compute the time evolved first and second moments of the quadrature operators. From the quadrature operator moments, 
we construct the covariance matrix which is used to measure the coherence in the system. The coherence is measured 
using the relative entropy of coherence measure. We consider three different classes of spectral densities in our analysis {\it viz}, 
the Ohmic, the sub-Ohmic, and the super-Ohmic densities. In our work, we study the dynamics of the coherent state, squeezed state, 
and displaced squeezed state. For all these states we observe that when the coupling with the system and the environment
is weak, the coherence monotonically decreases and eventually vanishes in a long time. Thus all the states exhibit Markovian
evolution in the weak coupling limit. In the strong coupling limit, the dynamics for the initial period is Markovian
and after a certain period, it becomes non-Markovian where we observe an environmental backaction on the system.  Thus
in the strong coupling limit, we observe a dynamical crossover from Markovian nature to non-Markovian behavior. This crossover
is very abrupt under some environmental conditions and for some parameters of the quantum state. Using a quantum master
equation approach we verify the crossover from the dynamics of the dissipation and fluctuation parameters and the results
endorse those obtained from coherence dynamics.
\end{abstract}

\maketitle

\section{Introduction}
Quantum coherence arises due to the superposition principle and is an essential property of quantum systems.
Basic features of coherence were well studied through \cite{glauber1963coherent}, where they were investigated
in the context of quantum optics. But these studies have focused only on the detection of quantum coherence
while a method to estimate coherence was not available. A procedure to measure coherence using a quantum
information theoretic framework was introduced by Baumgratz et. al. in Ref.~\cite{baumgratz2014quantifying}.
This method is used to estimate the coherence of finite dimensional systems.  This led to some fundamental results in the field of resource
theory of quantum coherence \cite{winter2016operational,brandao2015reversible,streltsov2017colloquium,chitambar2019quantum,chitambar2015relating}.
Subsequently quantum coherence has been studied in  the contexts, where the system is either in a
non-inertial frame of reference \cite{wang2016irreversible,he2018multipartite} or where the system is in contact
with an external environment \cite{Pati2018Banerjee,chandra2019time,radhakrishnan2019dynamics,cao2020fragility}.
But most of these investigations have been carried out on quantum systems with finite number of degrees
of freedom like qubits.

Quantum communication requires a quantum description of the interaction and propagation of electromagnetic waves.  An infinite number of
degrees of freedom with continuous spectrum is needed to describe the electromagnetic waves. Hence from both theoretical and experimental
perspective, continuous variable (CV) states \cite{braunstein2005quantum,adesso2014continuous,weedbrook2012gaussian,wang2007quantum}
of infinite dimensional systems are very important. Continuous variable systems constitute an extremely powerful resource to quantum
information processing. One particular class of continuous variable states is the Gaussian states
\cite{weedbrook2012gaussian,olivares2012quantum,wang2007quantum} for which the quantum state has a representation in terms
of Gaussian functions. To a first approximation, the ground states of the thermal states of a quantum system is a
Gaussian state. Similarly there are dynamical operations in which quantum
states are transformed to another Gaussian states and these are referred to as Gaussian operations. The Gaussian operations are linear in
nature and any nonlinear operation can be approximated to a Gaussian operation to a satisfactory level of accuracy. In light of these reasons,
the Gaussian states occupy a special place in the study of continuous variable systems. The first quantum resource to be studied extensively
in a Gaussian state is its entanglement. Since a covariance matrix along with the displacement vectors completely characterizes a Gaussian
state,  the entanglement can be quantified using a covariance matrix based approach. To measure the quantum coherence of Gaussian states
a measure based on relative entropy, using covariance matrix and displacement vectors was introduced in
Ref.~\cite{xu2016quantifying}. The fundamental postulates of {\it (i)} positivity {\it (ii)} monotonicity and {\it (iii)} convexity
were verified for this relative entropy measure.

In most of the investigations on discrete and continuous variable systems, we do not consider the effect of the
external environment on the system. But in reality the quantum systems are exposed to an environment, which acts as a bath
and affects the characteristics of the system by constantly interacting with it. To understand the effects of the environment and
incorporate them in the characterization of the quantum resources, we take an open quantum system approach. Here we model
the environment as a quantum many-body object which is coupled to the system. Based on the strength of the coupling and the
other characteristics of the system and the environment, there is a dynamical change in the quantum resource. Depending on
the bath spectrum and whether the system is weakly or strongly coupled to the environment, the system can exhibit a Markovian or
non-Markovian dynamics \cite{breuer2016colloquium,de2017dynamics}. A quantum resource can completely disappear and then
reappear a phenomenon known as revival. These features are dependent not only on the initial properties of the bath and the system
but also on how they are coupled to each other. Understanding this dynamical change is essential to the fabrication of quantum devices
\cite{shnirman2003noise,astafiev2006temperature,yoshihara2006decoherence,tong2006signatures,kakuyanagi2007dephasing,
seoanez2007dissipation,ribeiro2015non,vyas2020master}.
It  provides us a clear idea of the operational time of the quantum devices, i.e., the time within which the quantum operations have to
be completed before the quantum resources vanish.  The time evolution of the entanglement of finite dimensional systems has been
investigated in a very detailed manner and some unique features like sudden death of entanglement
\cite{yu2004finite, yu2006sudden,yu2009sudden} and entanglement revival
\cite{bellomo2007non,deng2021sudden,xu2010experimental,mazzola2009sudden}
due to environmental back action have been observed. For continuous variable systems, investigations on the dynamics of
entanglement have been done  \cite{Liu2007Goan,Zhang2007hong,Paz2008dynamics}. An open quantum system study of coherence
has been carried out using an atom-field interaction model \cite{chandra2019time} and a central spin model
\cite{radhakrishnan2019dynamics}. The dynamics of coherence and its distribution was extensively  investigated for the different
bipartite and tripartite states. But both these investigations were carried out on finite dimensional system. For the continuous variable
systems, the quantum coherence dynamics has not been investigated so far. In this work we investigate the dynamics of coherence
of a single mode Gaussian states {\it viz} the coherent state, the squeezed state and the displaced squeezed state.

The plan of the manuscript is as follows:  In Sec.~\ref{states and coherence} we give a brief introduction to the continuous variable
quantum state and explain the Gaussian states in detail.  Also we describe the covariance matrix based coherence measure introduced
in Ref.~\cite{xu2016quantifying}. The description of the Gaussian state in contact with a non-Markovian environment is
described in Sec.~{\ref{singlemodeNME}.
For the single mode coherent states the dynamics of quantum coherence is discussed in Sec.~\ref{coherentstates}. Next we
study the effect of non-Markovian environment on squeezed states in Sec.~\ref{squeezedstates}.  An analysis of coherence in
the single mode displaced squeezed state is shown in Sec.~\ref{displacedsqueezedstate}.  Finally in Sec.~\ref{crossovertransition},
we investigate the crossover from Markovian to non-Markovian dynamics.  We present our conclusions in Sec. \ref{conclusions}.

\section{Gaussian States and Quantification of coherence}
\label{states and coherence}

\subsection{Gaussian states}
A  continuous variable quantum system \cite{braunstein2005quantum,adesso2014continuous} has an infinite dimensional
Hilbert space in which observables have a continuous eigenspectrum.
An example of a continuous-variable system is the electromagnetic field in which the quantized radiation  modes are represented by the
bosonic modes. The tensor product Hilbert space corresponding to these modes is $\otimes_{k=1}^{n} H_k$,  where `$n$' is the
number of bosonic modes.  For a given mode `$k$' of the bosonic field, the operators  $a_k^{\dagger}$ and $a_k$ are the corresponding
creation and annihilation operators.  Apart from the bosonic field operators, the continuous-variable system can also be described using
the quadrature operators $\{x_k,p_k\}$.  The quadrature field operators can be arranged in the form of a $2n$-dimensional vector as
${\bm \xi} = \left(x_1, p_1,...,x_n, p_n\right)$.  Here the $2n$-dimensional vectors ${\bm \xi}$ contains the quadrature pairs of all the $n$ modes.
In terms of the bosonic field operators, the quadrature operators $\{x_k,p_k\}$ are expressed as
\begin{align*}
x_k = (a_k + a_k^{\dagger}),  \qquad  p_k = -i (a_k - a_k^{\dagger}),
\end{align*}
and they are canonically conjugate to each other. These quadrature field operators act similar to the canonically conjugate position and
momentum operators of a harmonic oscillator.

In the present work we study a single-mode continuous variable system.  The two quadrature operators corresponding to the single mode
continuous variable system are $\xi_1=x=(a + a^{\dagger})$  and $\xi_2=p=-i (a - a^{\dagger})$ and the 2D-vector
${\bm \xi} = \{\xi_{1}, \xi_{2} \}$. The quadrature operators satisfy the canonical commutation relations
$\left[ \xi_i, \xi_j\right]=2i \Omega_{ij}$, where $\Omega_{ij}$ are the elements of the matrix
\begin{align}
\bm{\Omega} = \left(\begin{array}{cc}
 0  & 1\\
-1 & 0\\
\end{array}
\right).
\end{align}
In particular, we study the coherence dynamics of the Gaussian states. The Gaussian states \cite{weedbrook2012gaussian,olivares2012quantum}
are defined as the states with Gaussian Wigner function. For a Gaussian state, the Wigner quasiprobability distribution is
\begin{align}
\nonumber
W({\bm \xi}) = \frac{1}{2\pi \sqrt{\rm{det}~\bm{V}}} \exp \left\{ - \frac{1}{2} ({\bm  \xi - \bm{\overline \xi} }).
\bm{V}^{-1} ({\bm  \xi - \bm{\overline \xi} })^{T}  \right\}
\label{wigner}
\end{align}
A Gaussian state is completely characterized by the first and the second moments of the quadrature field operators.  Here the
vector of the first moments ${\overline{\bm \xi}} = (\langle \xi_1 \rangle, \langle \xi_2 \rangle)$ and the covariance matrix
${\bm V}$
\begin{align}
V_{ij} = \langle \{ \Delta \xi_i, \Delta \xi_j \} \rangle = {\rm Tr} \left( \{ \Delta \xi_i, \Delta \xi_j \} \rho \right),
\end{align}
where $\Delta \xi_i=\xi_i - \langle \xi_i \rangle$ is the fluctuation operator and
$\{ \Delta \xi_i, \Delta \xi_j \} = (\Delta \xi_i \Delta \xi_j + \Delta \xi_j \Delta \xi_i )/2$.  For the single mode quantum system under
consideration, the matrix elements of the covariance matrix are
\begin{eqnarray}
V_{ii} &=& \langle \xi_i^2 \rangle - \langle \xi_i \rangle^2, \\
V_{ij} &=& \frac{1}{2} \langle \xi_i \xi_j + \xi_j \xi_i \rangle - \langle \xi_i \rangle \langle \xi_j \rangle.
\label{covele}
\end{eqnarray}
To reflect the positivity of the density matrix, the covariance matrix has to satisfy the uncertainty relation \cite{simon1994quantum}
$\bm{V} + i \bm{\Omega} \ge 0$.

\subsection{Measurement of Coherence}

The quantum coherence of a qubit system is measured as follows:  First we define the set of incoherent states $\mathcal{I}$,
i.e., the states with zero coherence. Next we use a suitable distance measure to find the distance between the
arbitrary density matrix and the closest incoherent states. In Ref.~\cite{baumgratz2014quantifying}, the authors used the
relative entropy measure to quantify the amount of coherence in the system. In general the relative entropy distance
between two density matrices is
\begin{align}
\mathcal{D}(\rho) =  \min_{\sigma} S(\rho \| \sigma) = \min_{\sigma} {\rm Tr} (\rho \log_{2} \rho - \rho \log_{2} \sigma),
\label{coher1}
\end{align}
where $\rho$ is the given quantum state and $\sigma$ is the reference state.  Using this formulation we can find the
coherence in the system by assuming the reference state to be the incoherent state.  It has been proved that for the relative
entropy of coherence, the closest incoherent state is $\rho_d=\sum_n \rho _{nn} \vert n \rangle \langle n \vert$, which is the
diagonal state of the density matrix. Hence it is not necessary to perform the minimization to determine the quantum coherence.
The relative entropy measure reads:
\begin{eqnarray}
C(\rho) =  S(\rho \| \rho_d) = S(\rho_{d}) - S(\rho),
\label{coher2}
\end{eqnarray}
where $S(\rho) = - {\rm tr} (\rho \log_{2} \rho)$ is the vonNeumann entropy of the system. So far, the relative entropy of coherence is the
most widely used measure. The quantum coherence of continuous variable system was first investigated in
\cite{zhang2016quantifying}. Here the authors used a relative entropy
based quantifier of coherence by considering the density matrix $\rho$ in the Fock basis. While this method is a valid process
to quantify coherence, it does not give a closed form expression for general Gaussian states. Alternatively in
Ref.~\cite{xu2016quantifying}, the coherence measure for a one-mode Gaussian state $\rho$ was defined as
\begin{align}
C(\rho) = \inf_{\delta}  S(\rho \| \delta).
\end{align}
Here $S(\rho \| \delta)$ is the relative entropy measure with $\delta$ being an incoherent Gaussian state
and the infimum runs over all incoherent Gaussian states. Also, it was proved that
a one mode Gaussian state is incoherent if and only if it is a thermal state. Hence for a Gaussian state, the minimization
is achieved for thermal states of the form
\begin{eqnarray}
\rho_d = {\overline \rho} = \sum_{n=0}^{\infty} \frac{\overline{n}^n}{(1+\overline{n})^{n+1}} \vert n \rangle \langle n \vert,
\label{thermal}
\end{eqnarray}
where $\overline{n}={\rm Tr} [a^\dagger a {\overline \rho}]$ is the mean number.  For the Gaussian states, the entropy of a
given state $\rho$  can be written  as
\begin{align}
S(\rho) = \frac{\nu+1}{2} \log_{2}  \frac{\nu+1}{2} - \frac{\nu-1}{2} \log_{2}  \frac{\nu-1}{2},
\end{align}
where $\nu = \sqrt{{\rm det} {\bm V}}$.  Based on these expressions, the relative entropy based coherence measure for the
Gaussian states is
\begin{eqnarray}
C(\rho) &=& S(\rho \| {\overline \rho}) ={ \rm Tr} (\rho \log_2 \rho - \rho \log_2 {\overline \rho}) \\
&=& \frac{\nu-1}{2} \log_2 \frac{\nu-1}{2} - \frac{\nu+1}{2} \log_2 \frac{\nu+1}{2} \nonumber \\
&+& (\overline{n}+1) \log_2 (\overline{n}+1) - \overline{n} \log_2 \overline{n}.
\label{coher3}
\end{eqnarray}
Hence the quantum coherence of a Gaussian state is completely determined by  the determinant of the covariance matrix
and the mean number $\overline{n}$ associated with the Gaussian state.  But this measure can be used to compute
coherence of only Gaussian states. It is important to note that when the system-environment interaction Hamiltonian
is bilinear in the creation and annihilation operators, the Gaussian nature of the quantum states is preserved
during the time evolution \cite{schumaker1986quantum,olivares2012quantum}.

\section{Single mode Gaussian states in a Non-Markovian environment}
\label{singlemodeNME}
A study of the dynamics of quantum coherence of qubit systems have been investigated both theoretically
\cite{Pati2018Banerjee,chandra2019time,radhakrishnan2019dynamics} and experimentally \cite{cao2020fragility,ding2021experimental}.
On the theoretical side, the dynamics was characterized in atomic systems and spin systems for both Markovian
and non-Markovian environments. Experimentally in \cite{cao2020fragility,ding2021experimental}, the dynamics of
coherence, entanglement and mutual information were compared in qubit system. In the present work we consider a
continuous variable system consisting  of a single bosonic mode of frequency $\omega_{0}$ coupled to a general
non-Markovian environment \cite{xiong2010exact,zhang2012general,ali2014exact,ali2015non,xiong2015non}
at finite temperature. The single bosonic mode can be either a quantum optical field, a superconducting resonator
or a nano-mechanical oscillator. A structured bosonic
reservoir with a collection of infinite modes of varying frequencies can describe a general non-Markovian environment. The entire
system comprising of the single bosonic mode and the reservoir can be described by the Fano-Anderson Hamiltonian
\cite{anderson1961localized,fano1961effects}. This Hamiltonian was introduced in the context of atomic  \cite{fano1961effects}
and condensed mater physics \cite{anderson1961localized} and has been used to study several different models.
The Fano-Anderson Hamiltonian of the system and bath combination is:
\begin{align}
H = \hbar \omega_0 a^{\dagger}a + \hbar \sum_k \omega_k b_k^{\dagger}b_k
      + \hbar \sum_k  \left(\mathcal{V}_k a^{\dagger} b_k  + \mathcal{V}_k^{\ast} b_k^{\dagger}a \right),
\label{Tot}
\end{align}
where $\omega_{0}$ is the system frequency and $a^{\dag}$ ($a$) is the creation (annihilation) operator
corresponding to the bosonic mode (system) and $b^{\dag}_{k}$ ($b_{k}$) is the creation (annihilation) operator of
the $k^{th}$ mode of the bosonic reservoir with frequency $\omega_{k}$.  The factor $\mathcal{V}_{k}$ represents
the coupling strength between the system and the environment.

The dynamics of the single mode system and the environment is solved using the Heisenberg equation of motion
approach.  The time evolution of the bosonic field operator and the  environment operators reads:
\begin{align}
a(t) = e^{\frac{i H t}{\hbar}} a e^{-\frac{i H t}{\hbar}},  \quad \quad  b_k(t) = e^{\frac{i H t}{\hbar}} b_k e^{-\frac{i H t}{\hbar}}.
\end{align}
In the Heisenberg picture, these operators satisfy the following equations of motion,
\begin{eqnarray}
&  & \frac{d}{dt} a(t) =  -\frac{i}{\hbar}  \left[ a(t), H \right] = -i \omega_0 a(t) - i \sum_k \mathcal{V}_k b_k(t),  \quad
\label{EOM1} \\
&  & \frac{d}{dt} b_k(t) = - \frac{i}{\hbar} \left[ b_k(t), H \right] = -i \omega_k b_k(t) - i \mathcal{V}_k^{\ast} a(t).   \quad
\label{EOM2}
\end{eqnarray}
Solving for Eqn. (\ref{EOM2}) for $b_{k}$ we get the solution
\begin{align}
b_k(t) = b_k(0) e^{-i\omega_k t} - i \mathcal{V}_k^{\ast} \int_0^t  d\tau~a(\tau)~e^{-i\omega_k(t-\tau)}.
\label{EOM3}
\end{align}
Substituting Eq. (\ref{EOM3}) in (\ref{EOM1}) we arrive at the following quantum Langevin equation \cite{ali2017nonequilibrium}
\begin{align}
{\dot a}(t) + i \omega_0 a(t)  + \int_0^t  d\tau g(t,\tau) a(\tau) = - i \sum_k \mathcal{V}_k b_k(0) e^{-i\omega_k t}.
\label{langevinequation}
\end{align}
Here the integral kernel $g(t,\tau)  = \sum_k |\mathcal{V}_k|^2 e^{-i\omega_k (t-\tau)}$ characterizes the
non-Markovian memory effects between the system and the environment.  For the continuous environment
spectrum $g(t,\tau) = \int_0^{\infty} d\omega J(\omega) e^{-i\omega(t-\tau)}$, where
$J(\omega)=\varrho(\omega) |\mathcal{V}(\omega)|^2$ is the spectral density which characterizes all the
non-Markovian memory of the environment on the system. Here $\varrho(\omega)$
being the density of states of the environment and $\omega$ is the continuously varying frequency of the bath.
Due to the linearity of Eq. (\ref{langevinequation}) the time evolved bosonic operator $a(t)$ can be expressed
\cite{de2000continuous} in terms of the initial field operators $a(0)$ and $b_{k}(0)$ of the system and the environment as
\begin{align}
a(t) = u(t) a(0) + f(t).
\label{fieldoperators}
\end{align}
The time-dependent coefficient $u(t)$ and noise operator $f(t)$ are determined by the quantum Langevin equation,
and they satisfy the two integrodifferential equations given below:
\begin{eqnarray}
\frac{d}{dt} u(t) &=&   - i \omega_0 u(t) - \int_0^t d\tau g(t,\tau) u(\tau),
\label{utime} \\
\frac{d}{dt} f(t)  &=&   -i \omega_0 f(t)  -  \int_0^t d\tau g(t,\tau) f(\tau) \nonumber  \\
                            &   & - i \sum_k \mathcal{V}_k b_k(0) e^{-i\omega_k t}.
\label{ftime}
\end{eqnarray}
To determine $u(t)$ we solve Eqn. (\ref{utime}) numerically with the initial condition $u(0) =1$.
Here the solution for the noise operator obtained using the initial condition $f(0) =0$ is
\begin{align}
f(t) = - i \sum_k \mathcal{V}_k b_k(0) \int_0^t d\tau e^{-i\omega_k \tau} u(t,\tau).
\label{ft2}
\end{align}
The nonequilibrium thermal fluctuation can be evaluated using the quantity
\begin{align}
\langle f^{\dagger}(t) f(t)\rangle = v(t) = \! \! \int_{0}^{t}  \! \! \! d\tau_1   \int_{0}^{t^\prime} \! \! \!   d\tau_2 u(t,  \tau_1)
{\widetilde g}(\tau_1,\tau_2) u^{\ast}(t^{\prime},  \tau_2),
\label{fdf}
\end{align}
where we consider the initial state of the total system to be $\rho_{tot}(0)=\rho_S(0) \otimes \rho_E(0)$.
Here the initial environment state  $\rho_E(0) = \exp(-\beta H_E)/{\rm Tr}[\exp(-\beta H_E)]$ is the thermal state
for the Hamiltonian $H_E=\sum_k \hbar \omega_k b_k^{\dagger}b_k$, where $\beta = 1/k_{B} T$ is the inverse
temperature and $k_B$ is the Boltzmann constant.

When the environment has a continuous spectrum, the time correlation function reads:
\begin{align}
{\widetilde g}(\tau_1,\tau_2)  = \int_0^{\infty} d\omega  J(\omega) {\bar n}(\omega) e^{-i\omega(\tau_1-\tau_2)},
\end{align}
where ${\bar n}(\omega)=1/(e^{\hbar \omega / k_{B} T}-1)$ is the initial particle number distribution of the
bosonic environment.  From the Eqs. (\ref{fieldoperators}) to (\ref{fdf}), one can obtain the time-dependent
average values of the system operators as follows:
\begin{align}
\label{avg1}
&\langle a(t) \rangle = u(t) \langle a(0) \rangle, ~~\langle a^{\dagger}(t) \rangle = u^{\ast}(t) \langle a^{\dagger}(0) \rangle,  \\
\label{avg2}
&\langle a(t) a(t) \rangle = (u(t))^{2}  \langle a(0) a(0) \rangle, \\
\label{avg3}
&\langle a^{\dagger}(t) a^{\dagger}(t)\rangle = (u^{\ast}(t))^{2}  \langle a^{\dagger}(0) a^{\dagger}(0) \rangle, \\
\label{avg4}
&\langle a^{\dagger}(t) a(t) \rangle =  |u(t)|^{2}  \langle a^{\dagger}(0) a(0) \rangle + v(t).
\end{align}
Here we consider $\langle f(t)\rangle=\langle f^{\dagger}(t)\rangle=0$ and also
$\langle f(t) f(t) \rangle=\langle f^{\dagger}(t) f^{\dagger}(t)\rangle=0$, since the reservoir is initially in a thermal state
uncorrelated to the system.  Using the time-dependent average values in Eqs (\ref{avg1}) -(\ref{avg4}), we can
evaluate the time evolved first and second moments of the quadrature operators namely,
$\langle \xi_1(t) \rangle$, $\langle \xi_2(t) \rangle$, $\langle \xi_1^2(t) \rangle$,
$\langle \xi_2^2(t) \rangle$, $\langle \xi_1(t) \xi_2(t) \rangle$, and $\langle \xi_2(t) \xi_1(t) \rangle$.
The elements of the time evolved covariance matrix are
\begin{eqnarray}
V_{11} &=& 1+ 2 v(t) + 2  |u(t)|^{2} \,  {\rm Cov}(a^{\dagger}(0), a(0))  \nonumber  \\
                   &   &  + (u(t))^{2} \, {\rm Var}(a(0)) + (u^{\ast}(t))^{2} \, {\rm Var}(a^{\dagger}(0)), \qquad
\label{V11}
\end{eqnarray}
\begin{eqnarray}
V_{22} &=& 1+ 2 v(t) + 2  |u(t)|^{2} \,  {\rm Cov}(a^{\dagger}(0), a(0))  \nonumber  \\
                   &   &  - (u(t))^{2} \, {\rm Var}(a(0)) - (u^{\ast}(t))^{2} \, {\rm Var}(a^{\dagger}(0)), \qquad
\label{V22}
\end{eqnarray}
\begin{equation}
V_{12}  =  i  (u^{\ast}(t))^{2} \; {\rm Var}(a^{\dag} (0))  - i ((u(t))^{2}  {\rm Var}(a(0)).
\end{equation}
where ${\rm Cov}(a,b) = \langle a b \rangle - \langle a \rangle \langle b \rangle$ and ${\rm Var}(a) = {\rm Cov}(a,a)$ and
$V_{12} = V_{21}$ due to the symmetry of the covariance matrix.

Once the initial state and the bath parameters are known, the time evolved covariance matrix elements are completely
determined using the nonequilibrium Green's function $u(t)$ and $v(t)$. To calculate the Green's function we need to
specify the spectral density $J(\omega)$ of the environment. In our work we consider a Ohmic-type spectral density which
can simulate a large class of thermal baths \cite{leggett1987dynamics}
\begin{align}
J(\omega) = \eta ~\omega \left( \frac{\omega}{\omega_c} \right)^{s-1} ~e^{-\omega/\omega_c},
\label{sd}
\end{align}
where $\eta$ is the coupling strength between the system and the environment and $\omega_{c}$ is the frequency cut-off
of the environmental spectra. A localized mode is generated when the system-environment coupling approaches a
critical value $\eta_{c} = \omega_{0} /(\omega_{c} \Gamma(s))$ where $\Gamma(s)$ is the Gamma function.
Depending on the value of $s$, the environment is classified as Ohmic for $s=1$,
sub-Ohmic for $s<1$ and super-Ohmic for $s>1$. Throughout our work we use a scaled temperature
$T_{s} = k_{B} T/ \hbar \omega_{0}$, where $\omega_{0}$ is the system frequency. The coherence dynamics for
different environmental conditions, and various system-environment couplings is studied for pure Gaussian
states through our work.  Since the Hamiltonian under consideration (\ref{Tot}) is bilinear in the creation and annihilation operators,
during evolution, the Gaussian states preserve their form and remain Gaussian.

\begin{figure}
\includegraphics[width=\columnwidth]{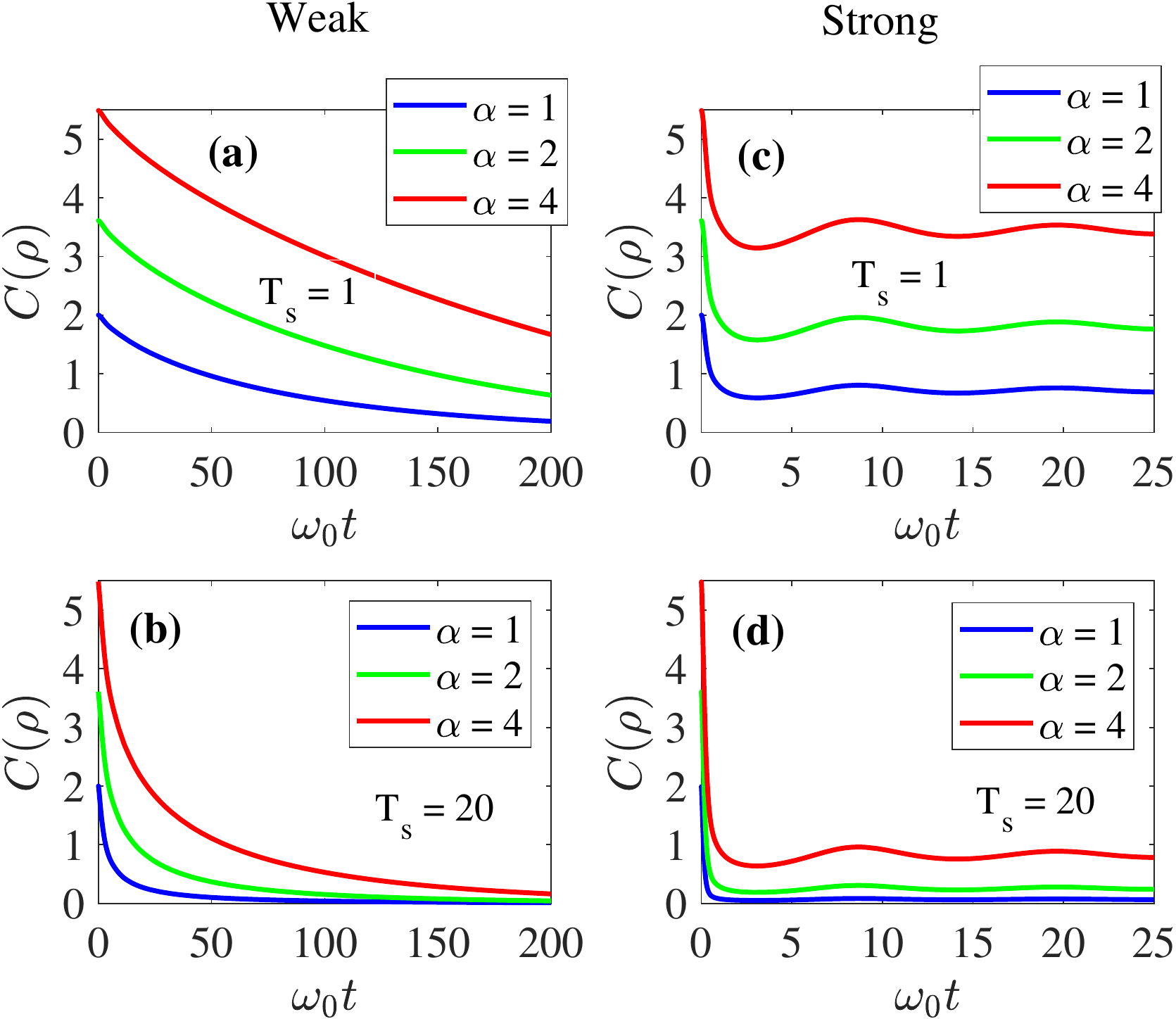}
\caption{The time evolution of quantum coherence of a coherent state with parameter $\alpha$ in contact with a Ohmic bath is shown for
weakly coupled systems ($\eta = 0.01 \; \eta_{c}$) in (a) at low temperature $T_{s} =1$, (b) at high temperature $T_{s} =20$ and
for the strongly coupled systems ($\eta = 2.0 \; \eta_{c}$) in (c) at low temperature  $T_{s} =1$, (d) at high temperature $T_{s} =20$.
We use the cut-off frequency $\omega_{c} = 5.0 \; \omega_{0}$. }
\label{fig1}
\end{figure}

\section{Quantum Coherence dynamics of coherent states in a Non-Markovian environment}
\label{coherentstates}
Glauber \cite{glauber1963coherent} defined a coherent state $| \alpha \rangle$ as the eigenstate of the annihilation operator $a$ with eigenvalue
$\alpha \in \mathbb{C}$.  We can generate the coherent state from the ground state $|0 \rangle$ of an oscillator through the action
of the displacement operator $ D(\alpha)  = \exp( \alpha a^{\dag}  - \alpha^{\ast} a)$, where $a^{\dag}$ and $a$ are the
annihilation and creation operators of the standard harmonic oscillator. In the number basis, the coherent states
can be expressed as
\begin{align}
|\alpha \rangle = \exp\left( - \frac{1}{2} |\alpha|^{2} \right)  \sum_{n=0}^{\infty}  \frac{\alpha^{n}}{\sqrt{n !}}  |n \rangle.
\end{align}
where $|n \rangle$ is the oscillator number basis. The coherent states are Gaussian wavepackets
which does not spread and has minimal uncertainty.  Thus for a free classical mode, the closest quantum analogue is the
coherent state.  Hence the coherent state is very useful in quantum optics especially to study the state of the quantized
electromagnetic field.  The time evolution of the quantum coherence of a coherent state subjected to a spectral density of the
form given in Eq. (\ref{sd}) is reported in this section. Depending on the value of the parameter $s$, the environments are
classified as Ohmic $s=1$,  sub-Ohmic $s<1$ and super-Ohmic $s>1$.

\noindent{\it Ohmic bath  (s=1):} \\
The time dynamics of quantum coherence of a coherent state for a Ohmic bath with spectral density
$J(\omega) = \eta \omega \exp(- \omega / \omega_{c})$ is shown in Fig. \ref{fig1}.  From the plots Figs \ref{fig1} (a) - (d) we find
that higher the value of the coherent state parameter `$\alpha$' higher the initial amount of coherence and also the
value at a given time.   In Fig  \ref{fig1}(a) and \ref{fig1}(b) the time variation of coherence, when the system is weakly coupled to the
environment i.e., $\eta =0.01 \, \eta_{c} $ is shown.  Particularly in Fig. \ref{fig1}(a) we look at the low temperature limit and in Fig. \ref{fig1}(b),
the high temperature regime of the coherence dynamics.  From both these plots we find that quantum coherence decreases
monotonically with time.  But it falls faster at higher temperature because apart from the environmental effects, the
thermal effects also contribute to the system decoherence.  The dynamics of coherence when the system is strongly coupled
to the environment i.e., for $\eta=2.0 \,  \eta_{c}$ is shown in Fig. \ref{fig1}(c) and \ref{fig1}(d). Here we can see that the coherence initially decreases
monotonically and then there is an oscillatory phase.  This oscillatory  phase is because of the environmental back action
due to the non-Markovian nature of the bath. For lower tempertures as shown in Fig. \ref{fig1}(c), the coherence saturates at a
higher value compared to the high temperature limit given in Fig. \ref{fig1}(d).

\begin{figure}
\includegraphics[width=\columnwidth]{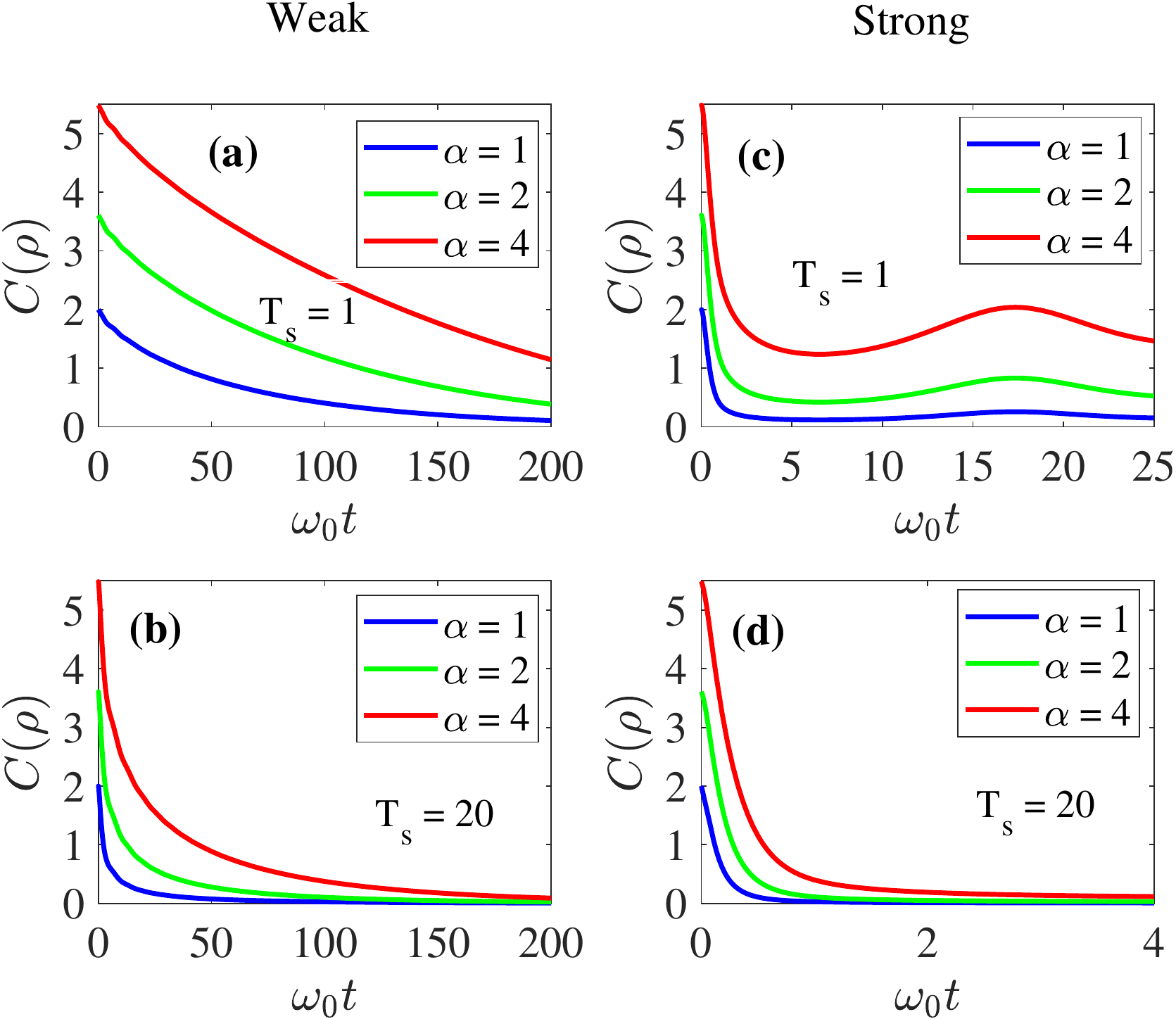}
\caption{The coherence dynamics of a coherent mode with parameter $r$ in contact with sub-Ohmic bath is studied for
weakly coupled systems ($\eta = 0.01 \; \eta_{c}$) in (a) low temperature $T_{s} =1$, (b) high temperature $T_{s} =20$.
For the strongly coupled systems ($\eta = 2.0 \; \eta_{c}$)  the plots are (c) at low temperature $T_{s}=1$ and (d) at high temperature $T_{s}=20$.
We use the cut-off frequency $\omega_{c} = 5.0 \; \omega_{0}$. }
\label{fig2}
\end{figure}

\noindent{\it Sub-Ohmic bath (s=1/2):}   \\
To investigate the coherence dynamics in the sub-Ohmic region we consider $s=1/2$ in Eq. (\ref{sd}).
The corresponding spectral density reads $J(\omega) = \eta \sqrt{\omega \, \omega_{c}} \exp(- \omega/ \omega_{c})$.
The results corresponding to this analysis is presented in Fig. \ref{fig2}.   The time variation of
coherence for the weak coupling limit  with $\eta =0.01 \, \eta_{c} $ is given in Fig. \ref{fig2} (a) and (b).  In Fig. \ref{fig2} (a) the
low temperature  limit is analyzed and we find that the coherence decreases monotonically with time.  At higher
temperatures also, coherence decreases monotonically but the fall is much faster as seen in Fig. \ref{fig2} (b).
This faster fall is because at high temperatures apart from the dissipation, the thermal fluctuations also cause
a decoherence in the system. The coherence dynamics when the system is strongly coupled to the environment $\eta=2.0 \,  \eta_{c}$ is
presented in Fig. (\ref{fig2}) (c) and (d) corresponding to the low and high temperature cases respectively.  In the
low temperature case illustrated in Fig. \ref{fig2} (c), we find a revival of coherence due to the non-Markovian effects
of the bath.  Such a coherence revival is not observed in the high temperature limit even in the strong coupling case.
This is because the temperature affects the environmental back action on the system, a feature which has also
been observed in Ref. [].

\begin{figure}
\includegraphics[width=\columnwidth]{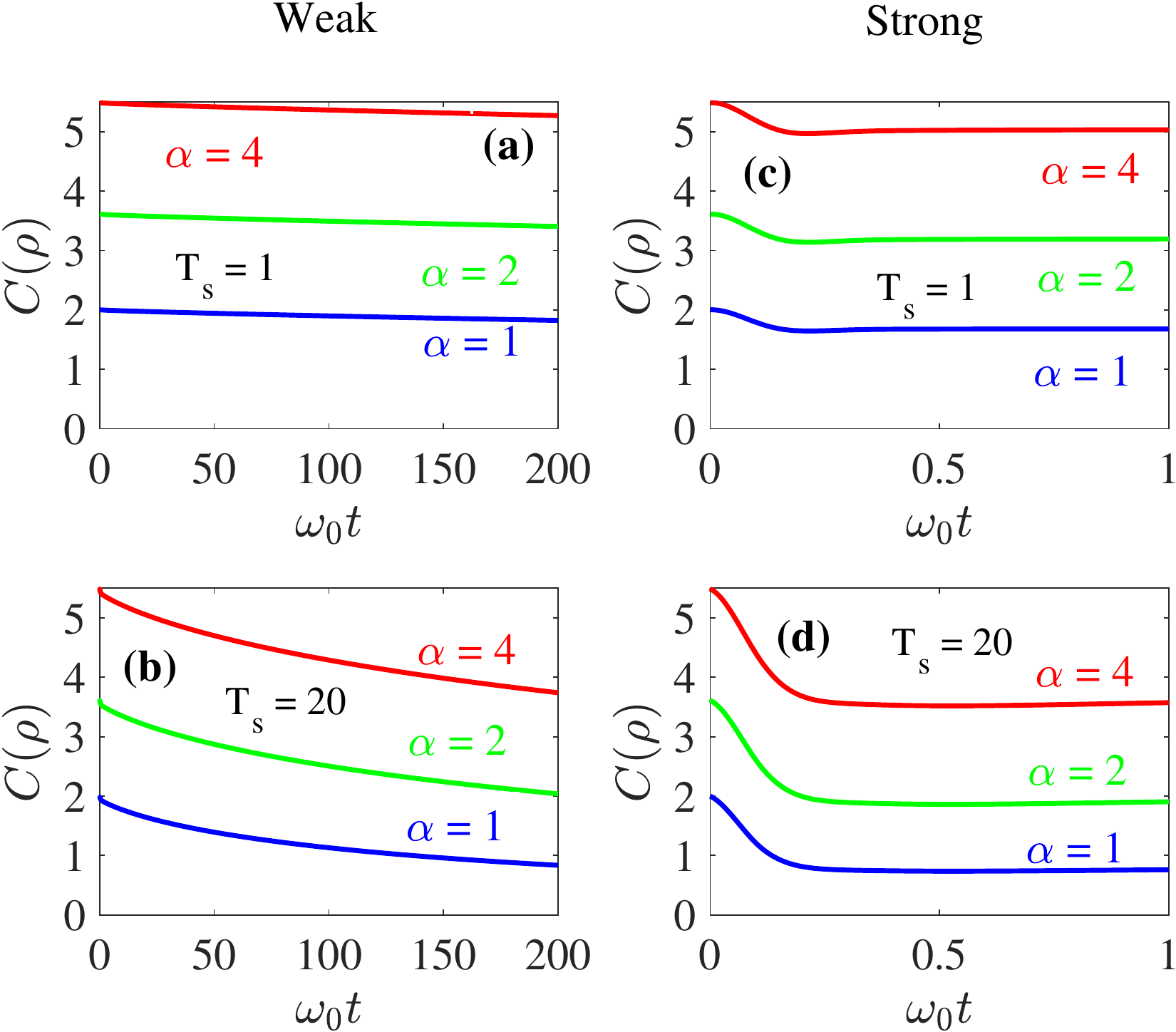}
\caption{A coherent mode in contact with a super-Ohmic bath is studied for weakly interacting systems ($\eta = 0.01 \; \eta_{c}$) for (a) $T_{s}=1$
and (b) $T_{s} = 20$ and also for strongly interacting systems ($\eta = 2.0 \; \eta_{c}$) with (c) $T_{s} =1$ and (d) $T_{s}=20$.  The cut-off
frequency used is $\omega_{c} = 5.0 \; \omega_{0}$.}
\label{fig3}
\end{figure}

\noindent{\it Super-Ohmic bath (s=3):}   \\
For the super-Ohmic bath we consider $s=3$ to investigate the dynamics of coherence. The spectral density in this case
reads $J(\omega) = \eta \, \omega_{c} (\omega / \omega_{c})^{3} \exp(- \omega/ \omega_{c})$. The coherence evolution
in this case is described through the plots in Fig. \ref{fig3} (a) to (d) considering different coupling strengths and different
temperatures.  In  Fig. \ref{fig3} (a) and (b) we analyze the coherence variation in the weak coupling limit when $\eta =0.01 \, \eta_{c} $.
Here we can see that the coherence decreases monotonically both in the low and high temperature limits.  Due to temperature
dependent decoherence effects, the rate of decrease is higher in Fig \ref{fig3} (b).   The nature of coherence variation in
the strong coupling limit ($\eta=2.0 \,  \eta_{c}$) is examined in Fig. \ref{fig3} (c) and (d) respectively.  In both these cases we see that
the coherence decreases initially and then saturates at a finite value a phenomenon known as coherence freezing.  But the
value at which coherence freezes is different in Fig. \ref{fig3} (c) and Fig. \ref{fig3} (d) and is also dependent on the parameter
$\alpha$.  For a higher $\alpha$, the coherence freezes at a higher value. The general decrease in coherence
is due to the environmental effects.  For the same environment at different temperatures, the system experiences
temperature dependent decoherence effects.

\begin{figure}
\includegraphics[width=\columnwidth]{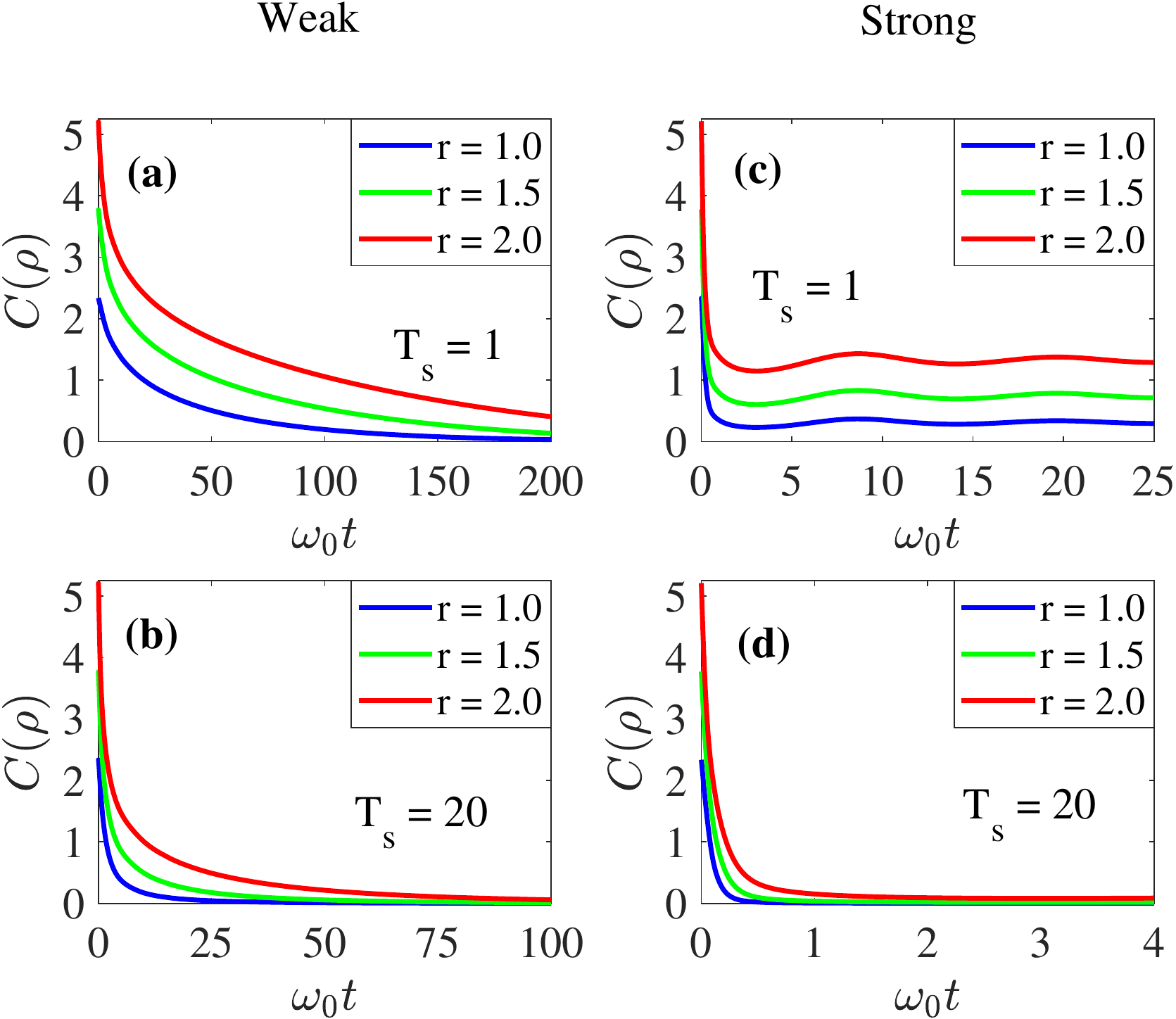}
\caption{A squeezed mode in contact with a Ohmic bath is studied for weakly interacting systems ($\eta = 0.01 \; \eta_{c}$) for (a) $T_{s}=1$
and $T_{s} =20$ and also for strongly coupled systems ($\eta = 2.0 \; \eta_{c}$) for  (a) $T_{s}=1$  and $T_{s} =20$.  The cut-off frequency used is
$\omega_{c} = 5.0 \, \omega_{0}$. }
\label{fig4}
\end{figure}

\section{Non-Markovian dynamics of Squeezed states}
\label{squeezedstates}
A squeezed state has a minimum value for the product of the dispersion of the position and momentum operators.
For a squeezed state \cite{walls1983squeezed}, the Hamiltonian consists of terms quadratic in the creation and the
annihilation operator. The Gaussian unitary corresponding to the one mode squeezing operator is
\begin{align*}
S(r) = \exp \left[ r \left(a^{2} -( a^{\dag})^{2} \right) \right].
\end{align*}
Here we assume $r \in \mathbb{R}$ for the sake of convenience.  The Bogoliubov transformation of the annihilation and creation
operator based on the squeezing operator reads:
\begin{align}
S^{\dag} (r) a S(r) = a \cosh r - a^{\dag} \sinh r,, \nonumber \\
S^{\dag}(r) a^{\dag} S(r) = a^{\dag} \cosh r  - a \sinh r, \nonumber
\end{align}
When this squeezing operator is applied to a vaccum state we can generate a squeezed vaccum state
\begin{align*}
|0,r \rangle = S(r) |0 \rangle =  \frac{1}{\sqrt{\cosh r}}  \sum_{n=0}^{\infty}  \frac{\sqrt{(2n)!}}{2^{n} n!}  \tanh^{n} r  |2n \rangle.
\end{align*}
In this state, the variance of one of the quadrature operator is below the quantum shot noise and hence it is called a
squeezed state.  To compensate the squeezing in one quadrature, there is an antisqueezing in the other quadrature.
The dynamics of quantum coherence of the single mode squeezed state is presented in this section, for the spectral
density Eq. (\ref{sd}), considering the Ohmic, sub-Ohmic and the super-Ohmic conditions.

\noindent{\it Ohmic bath  (s=1) :} \\
The Ohmic bath has a spectral density $J(\omega) = \eta \omega \exp(-\omega/\omega_{c})$ and the dynamics of quantum
coherence of a single mode squeezed state is given through the plots in Fig. (\ref{fig4}).  When the squeezed single mode
system is weakly coupled ($\eta =0.01 \, \eta_{c} $) to the environment, the coherence evolution is shown through Fig. \ref{fig4}(a) and
Fig. \ref{fig4}(b) for the high and the low temperature limits respectively.  We do not observe non-Markovian effects in both the limits.
But the coherence falls faster in the high temperature limit due to temperature dependent decoherence effects.  Under
the conditions when the system is strongly coupled to the environment ($\eta=2.0 \,  \eta_{c}$), the coherence evolution is shown in
Fig. \ref{fig4} (c) and Fig. \ref{fig4} (d) respectively.  We observe a very small non-Markovian effects in the low temperature
limit.  It is absent in the high temperature limit due to the thermal decoherence effects.

\begin{figure}
\includegraphics[width=\columnwidth]{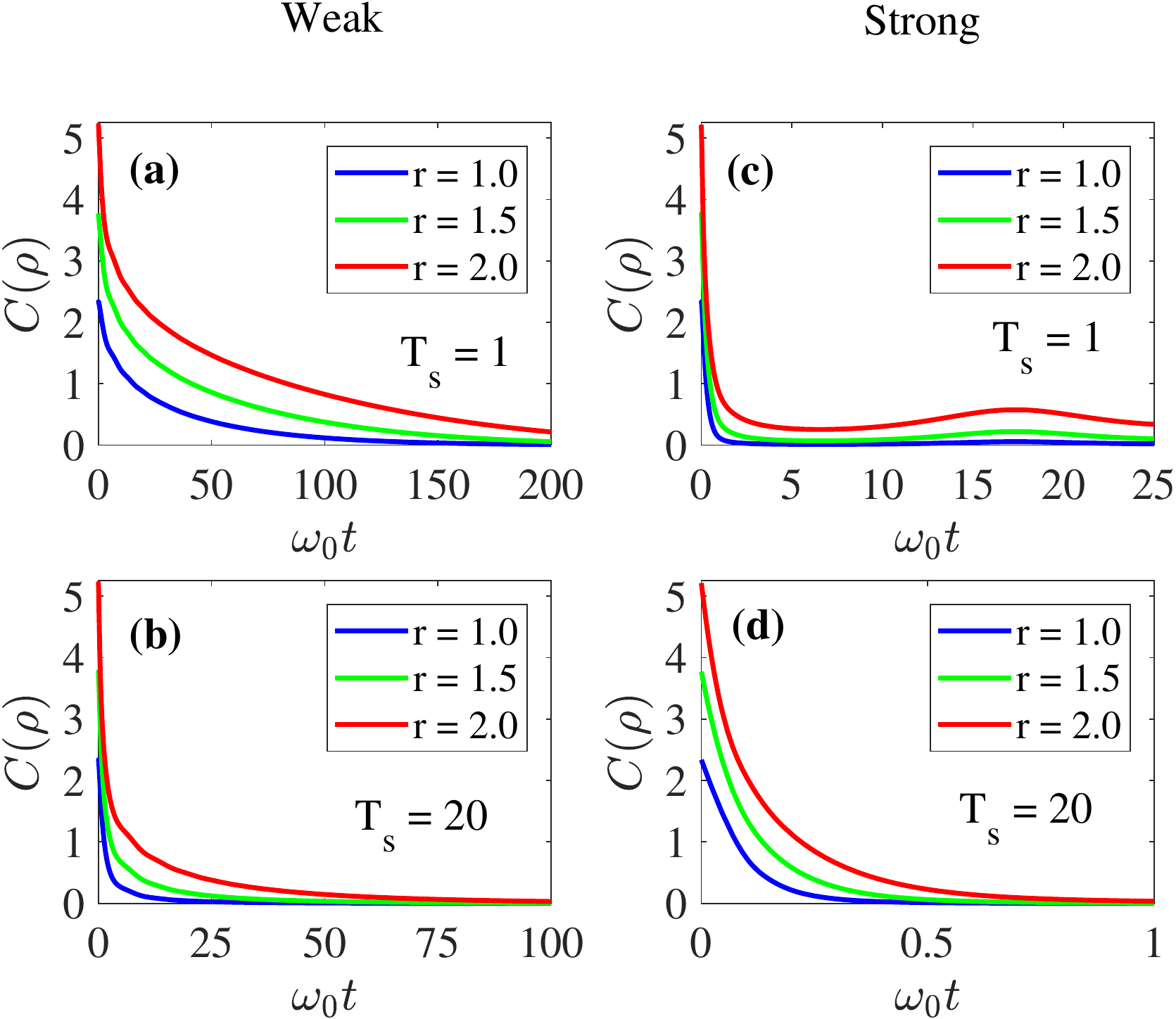}
\caption{The transient dynamics of coherence for a single squeezed mode in contact with a sub-Ohmic bath is studied for weakly
interacting systems ($\eta = 0.01 \; \eta_{c}$) for (a) $T_{s} =1$ and (b) $T_{s} =20$ and strongly interacting systems ($\eta = 2.0 \; \eta_{c}$)
(c) $T_{s} =1$ and (d) $T_{s}=20$ and using a cut-off frequency of $\omega_{c} = 5.0 \; \omega_{0}$. }
\label{fig5}
\end{figure}

\begin{figure}
\includegraphics[width=\columnwidth]{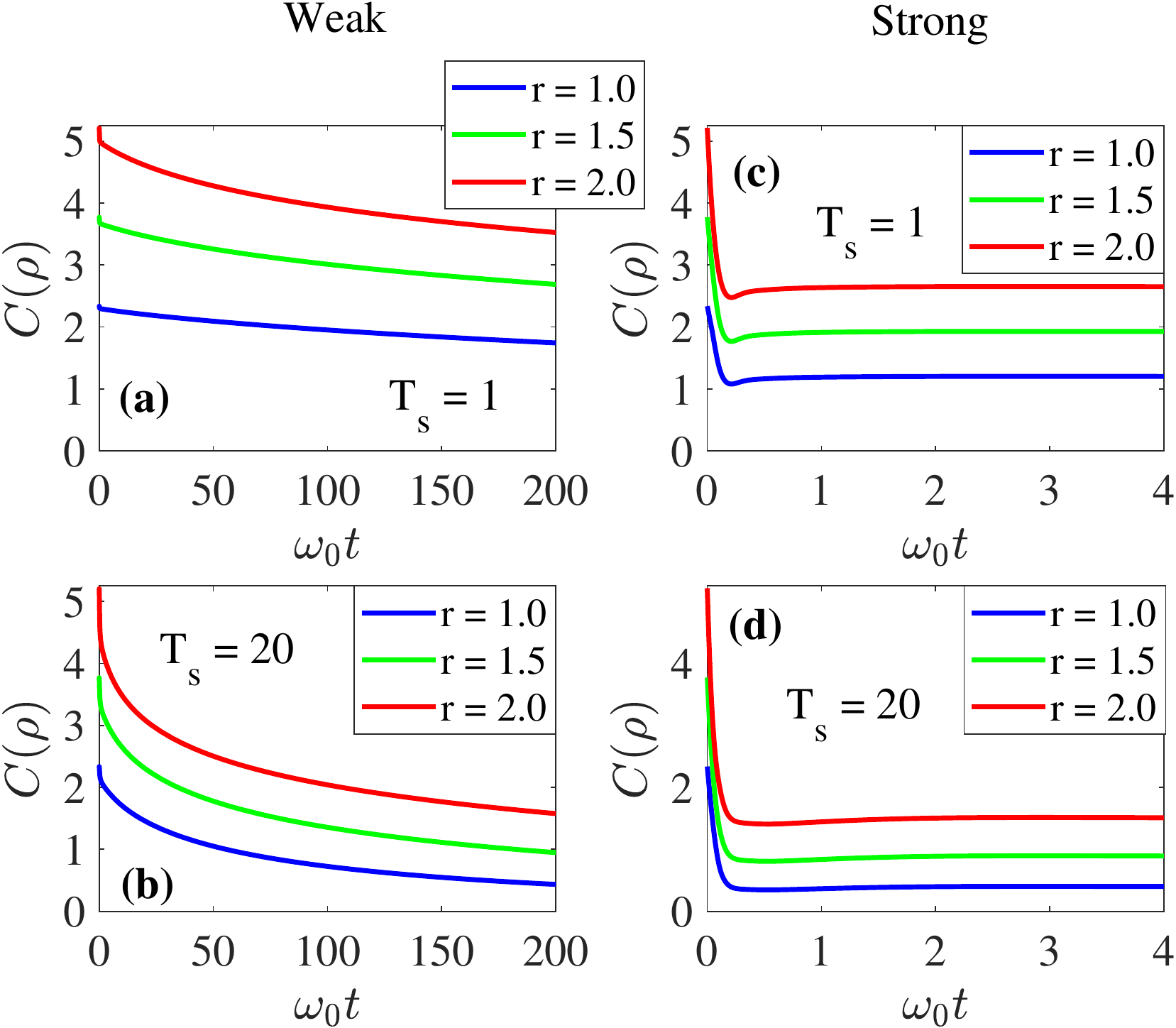}
\caption{For a squeezed mode in contact with a super-Ohmic bath, the coherence dynamics is studied for weakly interacting systems
($\eta = 0.01 \; \eta_{c}$) for (a) $T_{s} =1$ and (b) $T_{s} =20$ and strongly coupled systems ($\eta = 2.0 \; \eta_{c}$)
(c) $T_{s}=1$ and (d) $T_{s} =20$.  The cut-off frequency used is $\omega_{c} = 5.0 \; \omega_{0}$.  }
\label{fig6}
\end{figure}

\begin{figure*}
\includegraphics[scale=0.5]{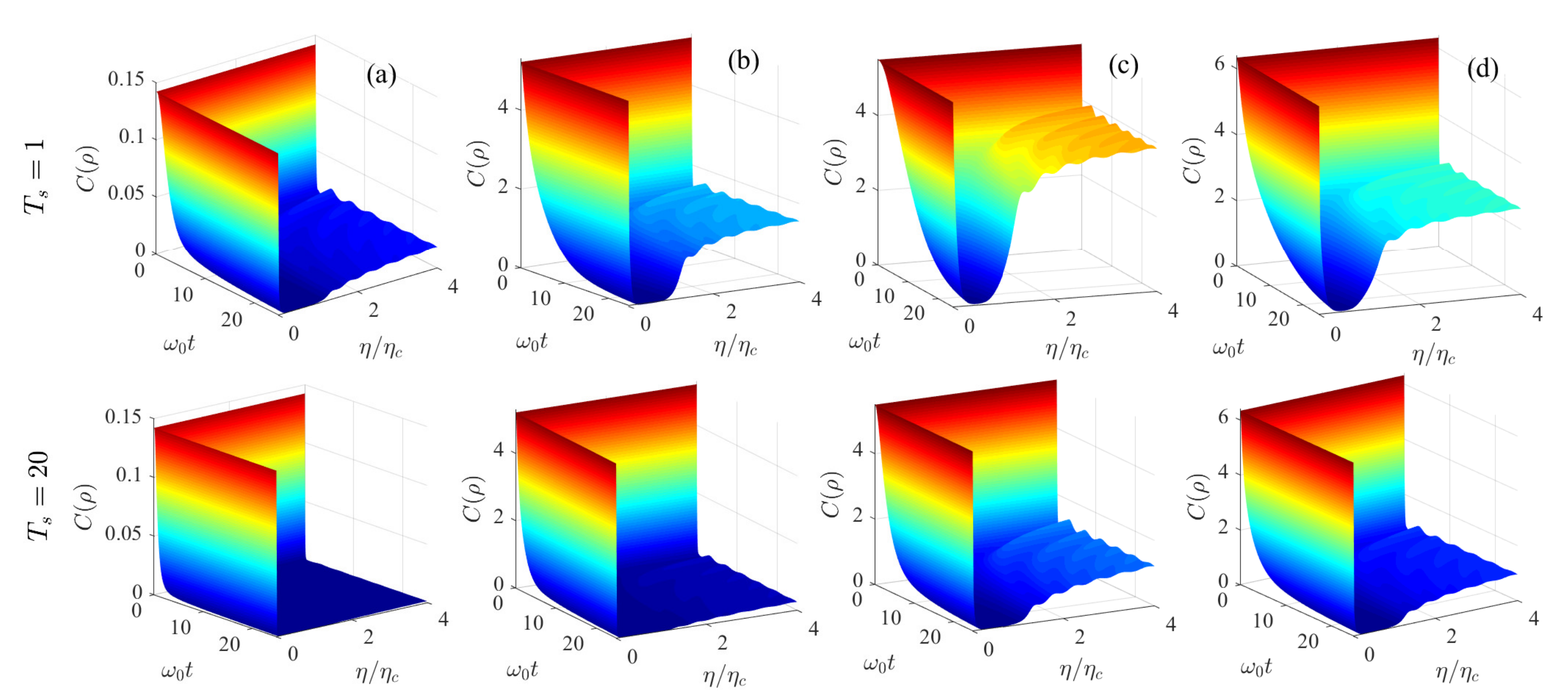}
\caption{The temporal evolution of the quantum coherence $(C)$ as a function of time ($\omega_{0} t$) and coupling strength
($\eta/\eta_{c}$) is given for the displaced squeezed state in contact with an Ohmic bath.  The plot is divided into four columns
as (a) $\alpha =0.1, r =0.1$, (b) $\alpha =0.1, r=2.0$ (c) $\alpha = 4.0, r=0.1$ and (d) $\alpha =4.0, r = 2.0$.  In each column
there are two plots one for $T_{s} = 1$ and $T_s =20$.  The cut-off frequency used is $\omega_{c} = 5.0 \; \omega_{0}$.  }
\label{fig7}
\end{figure*}

\noindent{\it Sub-Ohmic bath (s=1/2):}   \\
For a sub-Ohmic bath with spectral density, $J(\omega) = \eta \sqrt{\omega \, \omega_{c}} \exp(- \omega/ \omega_{c})$,
the coherence dynamics of the single mode squeezed state is shown via the plots in Fig. \ref{fig5} (a) - (d).  In Fig.
\ref{fig5} (a) and \ref{fig5}(b), the system is analyzed when it is weakly coupled to the bath ($\eta=0.01 \, \eta_{c} $).  Here we
see that at high temperature the coherence falls faster due to thermal decoherence. The strong coupling effects
($\eta=2.0 \,  \eta_{c}$) are analyzed in the plots \ref{fig5}(c) and \ref{fig5}(d) respectively.  In the low temperature limit, there is
a small non-Markovian effect, but in the high temperature limit, there is no environmental back action. Overall,
we find that coherence falls faster with increase in temperature.

\noindent{\it Super-Ohmic bath (s=3):}   \\
To study the effects of a super-Ohmic environment, we consider a bath with a spectral density
$J(\omega) = \eta \, \omega_{c} (\omega / \omega_{c})^{3} \exp(- \omega/ \omega_{c})$.  The plots in Fig. \ref{fig6}
(a) - (d) show the dynamics of the quantum coherence of the squeezed single mode state.  The change of coherence
with time, when the system is weakly coupled to the bath ($\eta =0.01 \, \eta_{c} $) is shown in Fig. \ref{fig6} (a) and (b).  The
coherence falls monotonically in the weak coupling limit and the rate of fall of coherence increases with increase in
the temperature.  In the strong coupling limit i.e., when ($\eta=2.0 \,  \eta_{c}$), the coherence initially decreases but saturates at a
finite value, a phenomenon known as coherence freezing.  The rate of fall of coherence and the saturation value of coherence
is dependent on the temperature and for higher temperature the coherence saturates at a lower value.
In general we find that the amount of coherence is directly proportional to the squeezing parameter $r$, under all the
environmental conditions.

\section{Quantum coherence evolution of displaced squeezed states}
\label{displacedsqueezedstate}

A  displaced squeezed \cite{kim1989properties} state has a general form as given below:
\begin{align*}
|\alpha, r \rangle = D(\alpha) S(r) |0\rangle,
\end{align*}
where $|0 \rangle$ is the vaccum state and $D(\alpha)$ and $S(r)$ are the displacement and squeezing operators
given below:
\begin{align*}
D(\alpha) = \exp(\alpha a^{\dag} - \alpha^{\ast} a);   \quad \quad  S(r) = \exp(r a^{2} - r (a^{\dag})^{2}).
\end{align*}
Here $\alpha \in \mathbb{C}$ and $r \in \mathbb{R}$ are the relevant parameters.  In this state, we investigate the transient dynamics
of quantum coherence for different values of the displacement parameter $\alpha$ and the squeezing parameter $r$.
The single mode displaced squeezed state is examined for the spectral density Eq. (\ref{sd}), considering the Ohmic,
sub-Ohmic and the super-Ohmic conditions.

\noindent{\it Ohmic bath  (s=1) :} \\
Ohmic bath has a spectral density  $J(\omega) = \eta \omega \exp(-\omega/\omega_{c})$ and the coherence dynamics
of the single mode displaced squeezed state for this spectrum is given through the $3D$ plots in Fig. \ref{fig7}.
The coherence $C(\rho)$ being the vertical axis and the parameters $\omega_{0}t$ and $\eta/\eta_{c}$ along the orthogonal
horizontal axes.  The figure is split  in to four columns labelled from (a) to (d) corresponding to the four different states of the displaced
squeezed states. The first row gives the plots in the low temperature limit ($T_{s} = 1$) and the second row the
high temperature regime ($T_{s} = 20$).

In the first column i.e., Fig. \ref{fig7} (a), we present the dynamics for the parameter values of ($\alpha =0.1, r =0.1$).
Both in the high and low temperature limit, the coherence falls monotonically to zero in a very fast manner which can
be referred to as coherence sudden death in analogy  with the sudden death of entanglement.  In the low temperature
limit the coherence revives and saturates at a finite value, which does not happen when the temperature is high.  After
revival in the low temperature limit, the coherence attains saturation and here it shows a non-Markovian feature that survives
for a long time.  When the squeezing is increased to high values ($\alpha=0.1, r=2.0$), the coherence behavior is shown in Fig. \ref{fig7} (b).
Here we again find a coherence sudden death in both the low and high temperature limits.  When the displacement
values are higher, the coherence is shown in Fig. \ref{fig7} (c) and \ref{fig7} (d) for the values ($\alpha=4.0, r=0.1$)
and ($\alpha=4.0, r=2.0$) respectively.  The system showns coherence sudden death after which there is a coherence
revival and saturation in this limit.  The system exhibits non-Markovian behavior for the parameters after the revival.
From all the plots Fig. \ref{fig7} (a) to (d) we can see that the saturation value is higher when ($\alpha=4.0, r=0.1$)
and on increasing the squeezing parameter it decreases as can be seen from Fig \ref{fig7} (d) corresponding to the parameter values
($\alpha=4.0, r=2.0$).

\begin{figure*}
\includegraphics[scale=0.5]{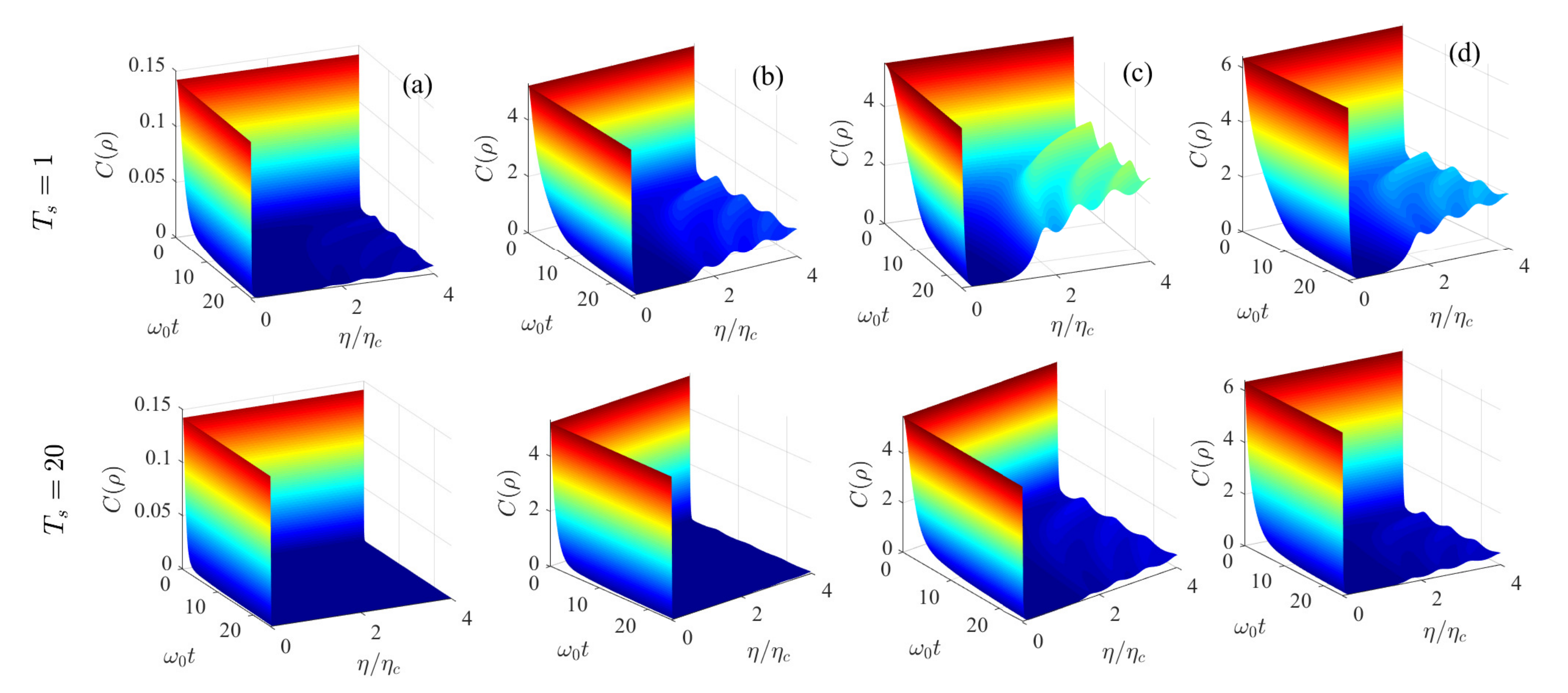}
\caption{The coherence dynamics of a displaced squeezed mode is shown in the figure above as a function of time
$\omega_{0}t$,  and $\eta/\eta_{c}$ when it is in contact with a sub-Ohmic bath.  The plot is divided into four columns
(a) $\alpha =0.1, r =0.1$, (b) $\alpha =0.1, r=2.0$, (c) $\alpha = 4.0, r=0.1$, and (d) $\alpha =4.0, r = 2.0$. In each
column there are two plots for $T_{s} =1$ and $T_{s} =20$. The cut-off frequency used is $\omega_{c} = 5.0 \; \omega_{0}$.}
\label{fig8}
\end{figure*}

\begin{figure*}
\includegraphics[scale=0.5]{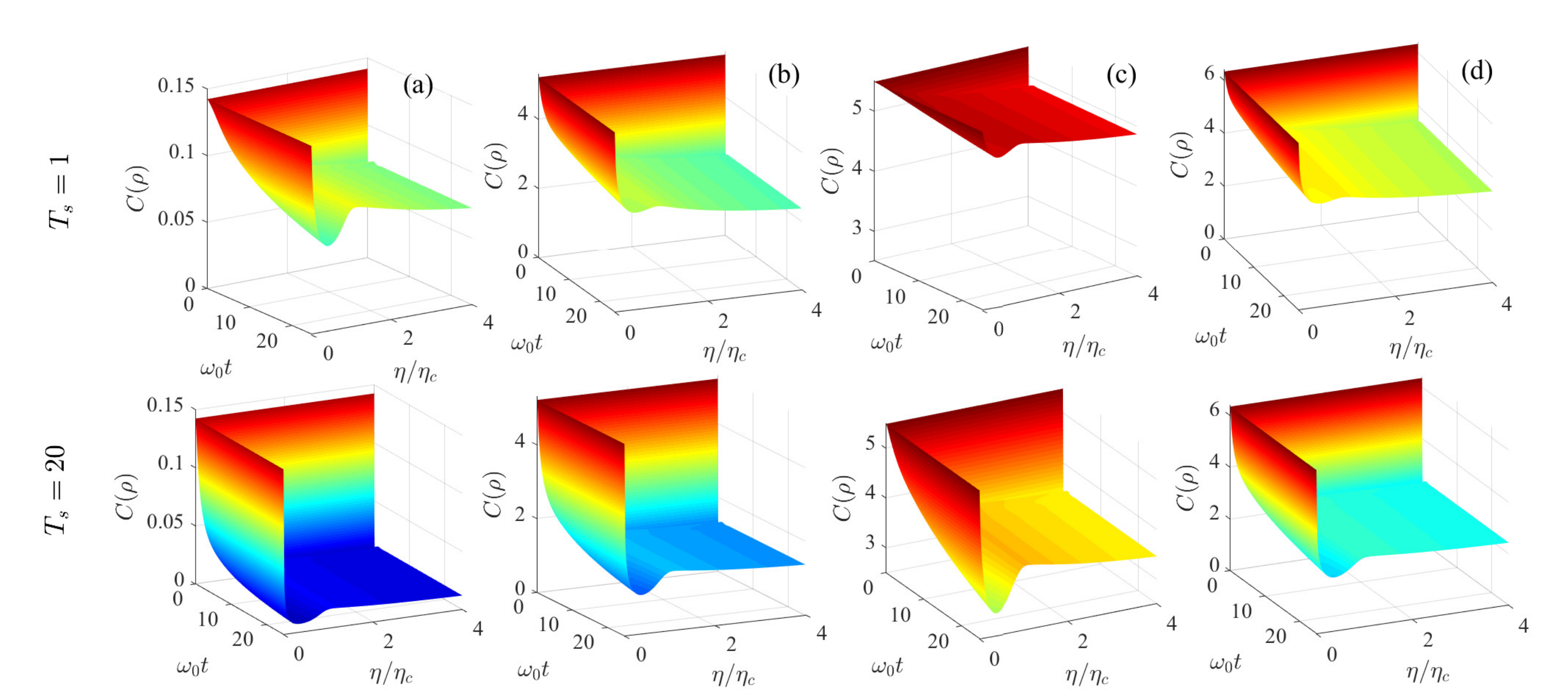}
\caption{The time evolution of coherence of a displaced squeezed mode in contact with a super-Ohmic bath is shown in
the figure above as a funciton of $\omega_{0}t$,  and $\eta/\eta_{c}$. There are four columns in the plot
(a) $\alpha =0.1, r =0.1$, (b) $\alpha =0.1, r=2.0$, (c) $\alpha = 4.0, r=0.1$, and (d) $\alpha =4.0, r = 2.0$.
In each column there are two plots for $T_{s} =1$ and $T_{s} =20$. The cut-off frequency used is
$\omega_{c} = 5.0 \; \omega_{0}$.}
\label{fig9}
\end{figure*}

\noindent{\it Sub-Ohmic bath (s=1/2):}   \\
The single displaced squeezed mode on being affected by a sub-Ohmic bath is given through the plots in Fig. \ref{fig8}.
The corresponding spectral density is $J(\omega) = \eta \sqrt{\omega \, \omega_{c}} \exp(- \omega/ \omega_{c})$.
The $3D$ plots describing the variation of coherence given as a set of plots, where the first row represents the
low temperature regime ($kT=1$) and the second row representing the high temperature regime ($kT=20$).
We label the columns from (a) to (d) in the plots.

For the parameter values of ($\alpha =0.1$,  $r=0.1$) and ($\alpha =0.1$,  $r=2.0$), the coherence dynamics
is given in Fig. \ref{fig8} (a) and (b) respectively.  Here in the low temperature limit we observe both coherence
sudden death and coherence revival and also a non-Markovian effect in the coherence obtained after the revival.
Meanwhile in the high temperature limit, we observe only coherence sudden death and there is no revival.
In the regimes ($\alpha =4.0$,  $r=0.1$) and ($\alpha =4.0$,  $r=2.0$) we  see the coherence again decaying
within a short interval of time as shown through the plots in Fig. \ref{fig8} (c) and (d) respectively.  The coherence sudden
death and the coherence revival is observed both in the low and high temperature limit.  Also the non-Markovian
effects are clearly seen in these plots.  In the sub-Ohmic limit, we see coherence revivals only in the high $\alpha$
region.

\begin{figure}
\includegraphics[width=\columnwidth]{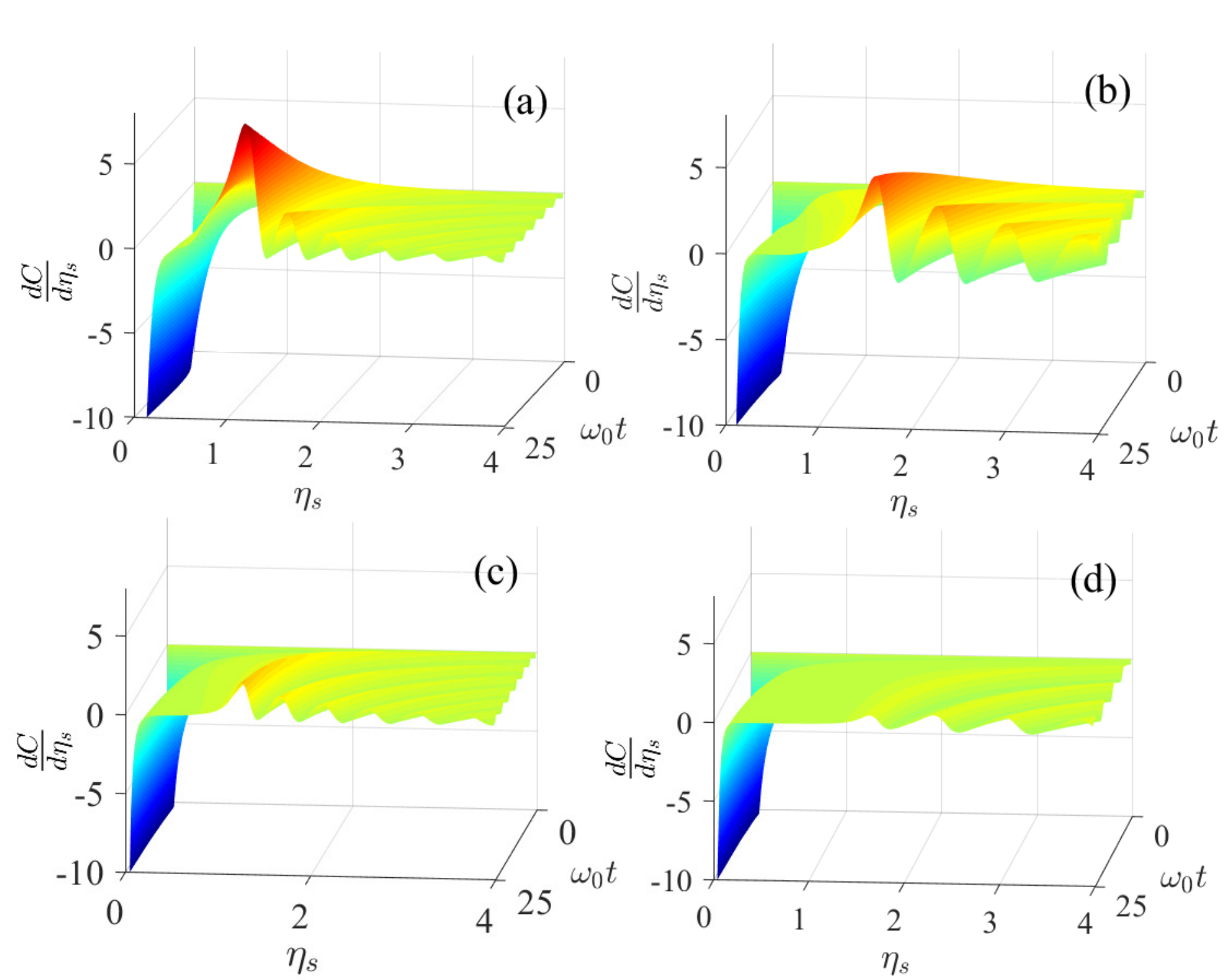}
\caption{The $3$-D plots $\frac{{\rm d} C}{{\rm d} \eta_{s}}$ Vs $\eta_{s}$ Vs $\omega_{0} t$ for the squeezed coherent
state with parameters $\alpha = 4.0, r = 0.1$ is given above.  In Fig. (a) and (c) we study the low temperature ($T_{s} =1$)
and the high temperature limit ($T_{s} = 20$) when the state is in contact with an Ohmic bath. The low temperature ($T_{s} =1$)
and high temperature limit ($T_{s} = 20$) is shown in Figs. (b) and (d) respectively, when the state is exposed to a  sub-Ohmic
bath. The cut-off frequency used is $\omega_{c} = 5.0 \; \omega_{0}$.}
\label{fig10}
\end{figure}

\begin{figure}
\includegraphics[width=\columnwidth]{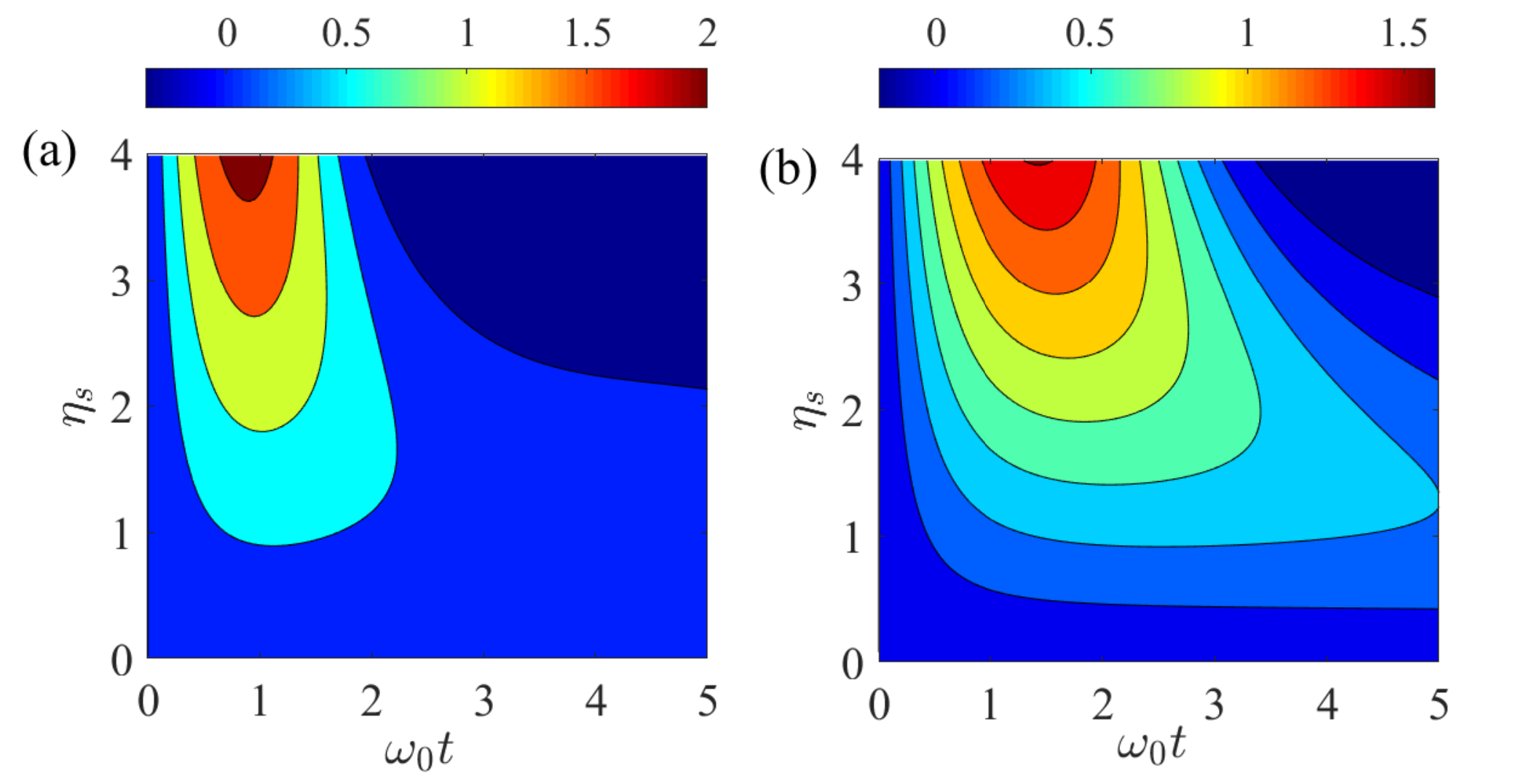}
\includegraphics[width=8.0 cm]{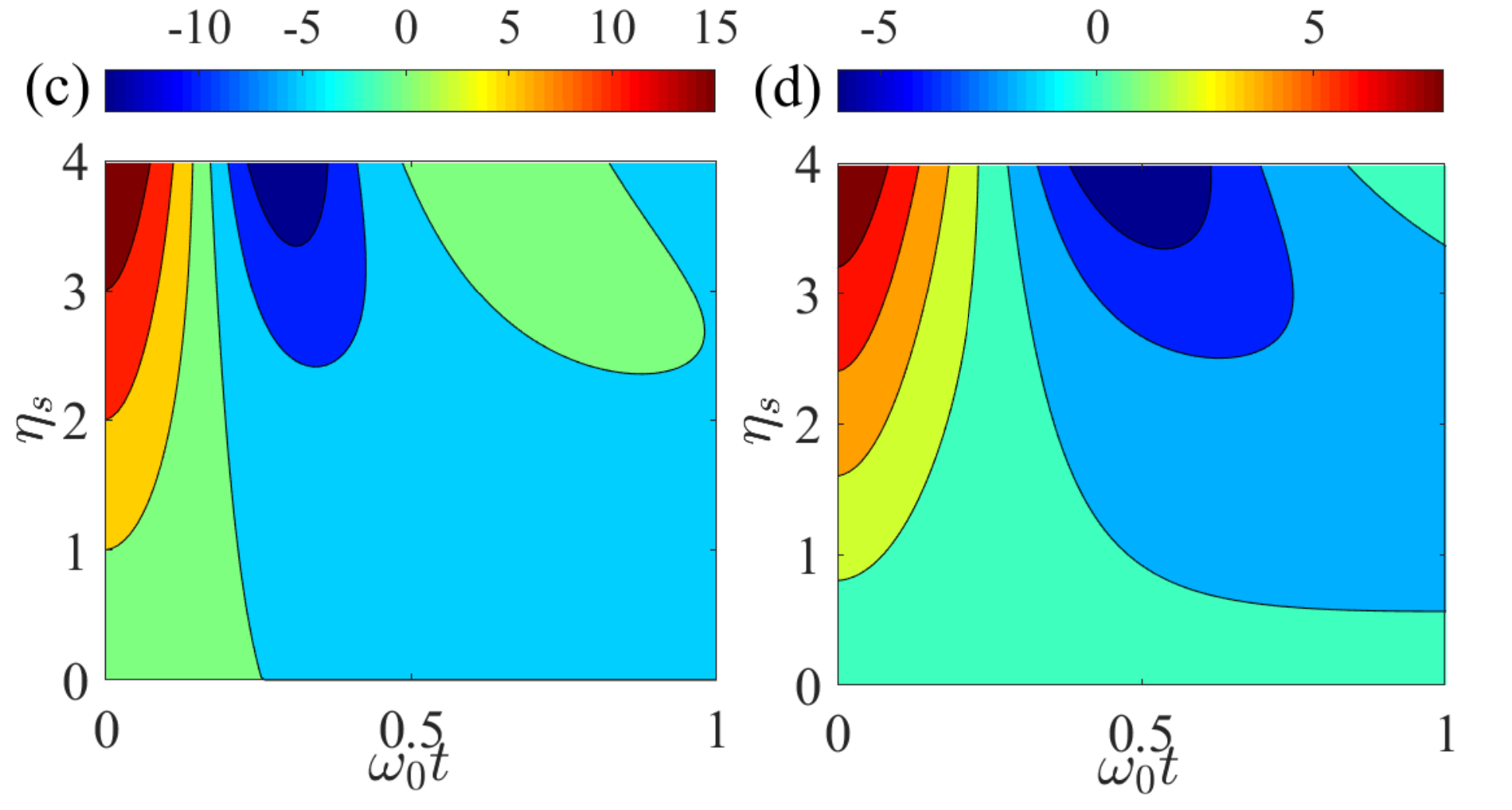}
\caption{The dissipation coefficient $\gamma(t)$ is calculated from the exact solution of $u(t)$.
We present here the contour plot of the dissipation coefficient $\gamma(t)$ and its time derivative
${\rm d}\gamma/{\rm d}t$ with varying time and coupling strength. In Fig. (a) and (b) we show
the transient behavior of $\gamma(t)$ for Ohmic ($s=1$) and sub-Ohmic ($s=1/2$) bath spectrum respectively.
We next plot the time derivative ${\rm d}\gamma/{\rm d}t$ with varying time and coupling strength
when the system is in contact with an Ohmic bath (c) and sub-Ohmic bath (d) respectively. The cut-off frequency
$\omega_c = 5.0 \omega_0$.}
\label{fig11}
\end{figure}

\noindent{\it Super-Ohmic bath (s=3):}   \\
The quantum coherence dynamics of the single displaced squeezed mode, when it is exposed to a super-Ohmic
environment is given here.  The spectral density used for this computation is
$J(\omega) = \eta \, \omega_{c} (\omega / \omega_{c})^{3} \exp(- \omega/ \omega_{c})$.  The
results pertaining to this study is given as $3D$ plots in Fig. \ref{fig9},  with the coherence $C(\rho)$ being along
the vertical axis and the parameters $\omega_{0} t$ and $\eta/\eta_{c}$ along the orthogonal horizontal axes.
The first row represents the low temperature regime ($T_{s}=1$) and the second row the high temperature regime
($T_{s}=20$) and the columns in the figure are labelled from (a) to (d).

\begin{figure}
\includegraphics[width=9.5cm]{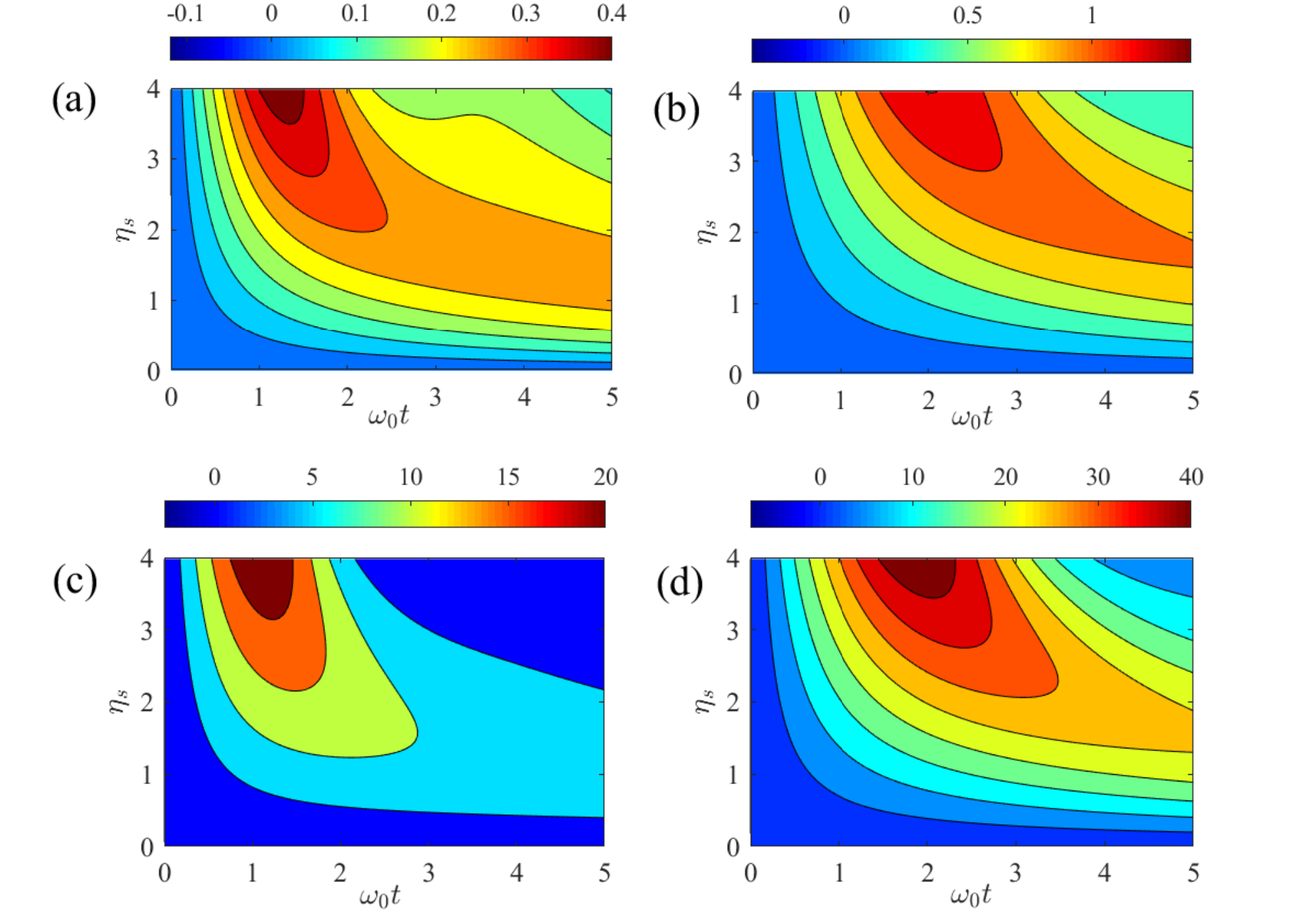}
\caption{The fluctuation coefficient $\widetilde{\gamma}(t)$ is calculated from
the exact solution of $u(t)$ and $v(t)$. We present here the contour plot of the
fluctuation coefficient $\widetilde{\gamma}(t)$ for Ohmic ($s=1$) spectrum in the
low temperature (a) $T_{s}=1$ and in the higher temperature (c) $T_{s}=20$
respectively. We next show the contour plot of fluctuation coefficient $\widetilde{\gamma}(t)$
for sub-Ohmic ($s=1/2$) spectrum in the low temperature (b) $T_{s}=1$ and in the high
temperature (d) $T_{s}=20$. The cut-off frequency in all the cases $\omega_c = 5.0 \omega_0$.}
\label{fig12}
\end{figure}

In Fig. {\ref{fig9}} (a) and (b) the coherence dynamics is given for parameter values of  ($\alpha =0.1$,  $r=0.1$)
and  ($\alpha =0.1$,  $r=2.0$) respectively.  For the parameter values ($\alpha =4.0$,  $r=0.1$) and
 ($\alpha =4.0$,  $r=2.0$), the dynamics is give through the plots in Fig. {\ref{fig9}} (c) and (d).  From all the
plots we observer that the coherence decays monotonically and then attains a saturation value at which the
coherence freezes for the rest of the evolution.  But the fall of coherence and the value at which the coherence
freezes is dependent on the temperature and higher the temperature, lower the saturation value.  This is
because apart from the loss of coherence due to dissipation, some of the coherence is also
lost due to the thermal decoherence.  From Fig. \ref{fig9} (a) we can see that initial coherence is low
when both the $\alpha$ and $r$ values are low.  On increasing either the displacement parameter $\alpha$
or the squeezing parameter $r$, we find that the system has a higher amount of initial coherence i.e.,
coherence at $t=0$ as shown in Fig. \ref{fig9} (b) and (c).  Further, from these two cases, we also observe
that the saturation value is higher when $\alpha$ is higher.  Finally in \ref{fig9} (d) we examine the situation
where both the displacement parameter ($\alpha$) and the squeezing parameter ($r$) are higher.
In this case we find that the initial amount of coherence is comparable to the cases in Fig. \ref{fig9} (b) and (c),
but the fall of coherence and the saturation value more closely resembles the case in  Fig. \ref{fig9}  (c),
where  ($\alpha =4.0$,  $r=0.1$).  An increase in either the displacement parameter or the squeezing parameter
increases the quantum coherence in the system.

\section{Markovian to non-Markovian crossover}
\label{crossovertransition}
The present work considers a Gaussian state in contact with an external non-Markovian environment.  When we
consider a Ohmic or sub-Ohmic bath coupled to the system, we initially observe a Markovian evolution for the coherent
state, squeezed state and displaced squeezed state. Depending on whether the coupling is weak or strong, the
dynamics continues to remain either Markovian or switches over to non-Markovian behavior.  In this current section
we examine the crossover behavior in the dynamics of a displaced squeezed state.

Let us consider the evolution of a displaced squeezed state with the parameter values of $\alpha = 4.0 $ and $r = 0.1 $.
When the coupling between the bath and the system is weak, we observe that the dynamics of the system is
completely Markovian  where the system experiences either a death of coherence or coherence freezing.  In
this case, the correlation time of the bath is much shorter than the system evolution time. The irreversible loss
of information occurs, because the bath states are restored very quickly and so any information received from
the state is lost. For systems which have a stronger coupling with the bath, the initial coherence decay is Markovian
and then there is a sudden switch where the dynamics becomes non-Markovian.  In the non-Markovian phase there
is a back flow of information from the environment to the system. To check this behaviour, we look at Fig. \ref{fig10},
the $3D$ plots of $\frac{{\rm d} C}{{\rm d} \eta_{s}}$ Vs $\eta_{s}$ Vs $\omega_{0} t$ in both the low temperature
($T_{s} = 1$) and the high temperature ($T_{s} = 20$) limits.  In Fig. \ref{fig10} (a), the plot shows ${\rm d} C / {\rm d} \eta_{s}$
for the Ohmic spectrum in the low temperature limit. Here we find that initially, the slope is $-ve$ and monotonic implying a Markovian nature.
Increasing the coupling strength, we observe that the slope changes from $-ve$ to $+ve$ indicating a change from
Markovian to non-Markovian behavior. For the Ohmic bath in the low temperature regime this crossover in the behavior
of dynamics is very abrupt and this sudden change is reminiscent of a phase transition.  In the initial phase we observe
$\tau_{b} \ll \tau_{s}$, where $\tau_{b}$ is the relaxation time of the bath and $\tau_{s}$ is the evolution time of the system.
After the crossover (i.e., transition) the time scales are related as $\tau_{b} \approx \tau_{s}$ indicative of a new phase.
In the low temperature limit $(T_{s} = 1)$, the sub-Ohmic regime is described in Fig. \ref{fig10} (b).  Here there is a region
where the coherence remains constant before the transition to the non-Markovian evolution.  The high temperature
$(T_{s} = 20)$ case of the $3$D plots are shown in Fig. \ref{fig10} (c) and (d) for the Ohmic and sub-Ohmic baths respectively.
From the plots we can observe that transition region between the Markovian and non-Markovian regimes is more pronounced.
Also the non-Markovian effect is decreased indicating a lesser amount of environmental back flow of coherence. This
reduced amount of back flow is due to the decoherence caused by thermal effects.

The dynamical behavior of the continuous variable system can also be studied using a quantum master equation.
The crossover between Markovian and non-Markovian can be perceived from the changing behavior of the
decay rates in the master equation. The total density matrix $\rho_{tot}$ describing the continuous variable system and
environment has a dynamics governed by the quantum evolution operator
$\rho_{tot}(t) = \exp\left( - \frac{i}{\hbar} H t \right)  \rho_{tot} (0) \exp\left( \frac{i}{\hbar} H t \right)$.
The initial state of the system is considered to be $\rho_{tot}(0) = \rho_{s} (0) \otimes \rho_{E} (0) $,
where $\rho_{E}(0) = \exp( - \beta H_{E}) / ({\rm Tr} \exp(- \beta H_{E}) )$  as proposed in
\cite{feynman1963Vernon,caldeira1983path}. Tracing over the environmental degrees of freedom we can get the
reduced density matrix of the system. The master equation for the reduced density matrix reads:
\begin{eqnarray}
\frac{{\rm d} \rho(t)}{{\rm d} t} &=& - i \omega_{0}^{\prime} (t) [ a^{\dag} a, \rho(t) ]  \nonumber  \\
                                                    &   &   + \gamma(t) [ 2 a \rho(t) a^{\dag}  - a^{\dag} a \rho(t) - \rho(t) a^{\dag} a ]  \nonumber \\
                                                    &   &   + \widetilde{\gamma} (t) [ a \rho(t) a^{\dag} + a^{\dag} \rho(t) a - a^{\dag} a \rho(t)
                                                                         - \rho(t) a a^{\dag} ]  \qquad
\end{eqnarray}
where the coefficients are
\begin{align*}
 \omega_{0}^{\prime} (t)  = {\rm Im} \left[ \frac{\dot{u}(t)}{u(t)} \right],   \qquad   \gamma(t) = - {\rm Re}  \left[ \frac{\dot{u}(t)}{u(t)} \right] \\
 \widetilde{\gamma}(t)  =  \dot{v}(t) - 2 v(t) {\rm Re}  \left[ \frac{\dot{u}(t)}{u(t)} \right].
\end{align*}
Here $\omega_{0}^{\prime} (t)$ is the renormalized frequency of the single mode system and $\gamma(t)$ and $\widetilde{\gamma}(t)$ are the
dissipation and fluctuation coefficients. In Fig.~\ref{fig11} (a) and (b), we present a contour plot of the dissipation coefficient
$\gamma(t)$ varying with time and coupling strength. The Ohmic case is described through the plot
in \ref{fig11}(a), where in the weak coupling limit $\eta_{s} \leq 1$  we find that the dissipation is almost a constant
throughout the entire evolution. For the strong coupling regime $\eta_{s} \geq 1$, the dissipation rate $\gamma(t)$ increases
initially and then decreases. This change in the direction of the decay rate with time, indicates a backflow of information from
the environment to the system.  A similar behavior of the dissipation constant $\gamma(t)$ is observed in the sub-Ohmic case
as shown in \ref{fig11}(b).  The reverse flow of information is more transparent from the contour plot of the slope of $\gamma(t)$.
We show contour plot of the time derivative of the decay rate (${\rm d}\gamma/{\rm d}t$) for the Ohmic case in Fig.~\ref{fig11} (c)
and sub-Ohmic case in Fig.~\ref{fig11} (d).  From the plot we observe that the slope of the dissipation coefficient $\gamma(t)$,
transitions from being positive to negative.  This is indicative of a crossover from the Markovian to a non-Markovian regime.

Next, we study the variation of the fluctuation coefficient namely $\widetilde{\gamma}(t)$ in Fig. \ref{fig12}.
We show the contour plot of fluctuation coefficient for Ohmic spectrum in the low temperature limit ($T_{s}=1$)
in Fig.~\ref{fig12} (a) and for the high temperature limit ($T_{s}=20$) in Fig. \ref{fig12} (c).  In the strong coupling
regime  ($\eta_{s} \geq 1$), the fluctuation coefficient $\widetilde{\gamma}(t)$ rises initially and then starts decreasing
with time.  This changing behavior of $\widetilde{\gamma}(t)$ indicates the environmental back action in the strong coupling
regime.  The contour plot of fluctuation coefficient for sub-Ohmic spectrum in the low temperature ($T_{s} = 1$)
is shown in Fig.~\ref{fig12} (b).  For the high temperature limit ($T_{s} = 20$) is given in Fig. \ref{fig12} (d).  We observe a
 similar dynamical crossover of $\widetilde{\gamma}(t)$ for the sub-Ohmic spectral density  as well.  Hence the
 environmental back flow of information \cite{breuer2009measure,liu2011experimental,chruscinski2014degree}
 i.e., the non-Markovian behavior can be obtained from the dynamical behavior of the dissipation and the fluctuation
 coefficients.  The dynamical crossover observed in the coherence dynamics of a single mode state,
 stands verified by the analysis carried out from the dynamics of the fluctuation and dissipation coefficients of the
 master equation of the state.

\section{Conclusion}
\label{conclusions}
The transient dynamics of quantum coherence of a single mode Gaussian state is analyzed in  the open quantum
system formalism.  For this we consider a non-Markovian environment, which is a collection of infinite bosonic modes
of varying frequencies.  The entire analysis is carried out in the finite temperature limit with arbitrary system-bath coupling
strength.  To measure the quantum coherence we use the relative entropy measure which estimates the distance to the
thermal state.  Hence the quantum coherence can be completely characterized by the determinant of the covariance
matrix and the average of the number operator of the Gaussian states.  We solve the quantum Langevin equation for the
bosonic mode operators to obtain the time evolved covariance matrix elements.  The time evolution of the field operators are
determined by the two basic nonequilibrium Green's functions namely $u(t)$ and $v(t)$.

The time dynamics of quantum coherence is studied for three distinct experimentally  realizable Gaussian states.  The
investigations have been carried out under three different environmental spectral densitites {\it viz} Ohmic, sub-Ohmic
and super-Ohmic spectral densitites.  When the system interacts weakly with the environment, the quantum coherence
decreases monotonically with time.  The coherence decay rate is relatively lower for the super-Ohmic bath when compared
with the Ohmic and the sub-Ohmic baths.  Also the rate of decay increases with the temperature.  In general we observe
a coherence death for the Ohmic and the sub-Ohmic baths in the short time limit.  For the super-Ohmic bath we observe coherence freezing
which is a saturation of coherence at a finite value.  In the strong coupling regime, the coherence decreases initially
and then starts increasing resulting in a revival of coherence for the Ohmic and the sub-Ohmic baths.  This is followed by
a stabilization of coherence with an oscillatory phase.  This oscillatory behavior is due to the environmental backaction
which is a feature of non-Markovian dynamics in the system.  Hence in our study we observe a transition or crossover
from the Markovian dynamics to the non-Markovian dynamics.  The non-Markovian memory effect helps us to restore
quantum resources in the strong coupling limit. The rate at which the crossover in the dynamics occurs
depends on the parameters of the Gaussian state as well as on the environmental parameters.  To verify the existence of the crossover
we study the dynamical behavior of the quantum system using a master equation approach.  Here we compute the dissipation and
fluctuation coefficients in the master equation of the reduced density matrix. Throughout the evolution, the
dissipation rate is almost a constant in the weak coupling limit.  In the strong coupling limit, the dissipation rate
increases initially and then decreases, which implies a Markovian to non-Markovian transition.  A  plot of the time
derivative of the decay rate shows that the slope of the dissipation coefficient changes from being positive to negative.
This is a clear indication of the crossover from the Markovian to the non-Markovian dynamics.  This crossover
feature is also verified from the dynamics of the fluctuation coefficient.  Thus we find that the quantum system has both
Markovian and non-Markovian behavior at different times and there is a dynamical crossover from the Markovian to
the non-Markovian regime.   This change from Markovian to non-Markovian regime is not a gradual change.
Instead it is either a sudden change or there is a time period between the Markovian and non-Markovian regimes,
where there is no dynamical change. This raises an important question as to whether this dynamical change can
be considered as some kind of phase transition between two different regimes with totally different relaxation time
scales.  While the results seems to point towards such a conclusion, a more detailed study from the point of view
of quantum phase transitions is needed to confirm and validate this, and such a study will form the scope of our future works.
A similar observation on an abrupt change from Markovian to non-Markovian dynamics has also been made in
Ref. \cite{pang2018abrupt}.  Here the authors study a single qubit system in contact with a single qubit environment.
In our work we consider a single mode Gaussian state in contact with a structured bosonic reservoir with a collection of
infinite modes to describe a general non-Markovian environment.  The investigations carried out in the present work can be
experimentally verified by measuring the quantum coherence of Gaussian systems
\cite{bowen2004experimental,furusawa1998unconditional,yonezawa2004demonstration,diguglielmo2007experimental}.

\section*{Acknowledgements}
Md.~Manirul Ali was supported in part by the Centre for Quantum Science and Technology, Chennai Institute of Technology, India,
vide funding number  CIT/CQST/2021/RD-007.
Chandrashekar Radhakrishnan was supported in part by a seed grant from IIT Madras to the Centre for
Quantum Information, Communication and Computing.


%

\end{document}